\documentclass[12pt,a4paper]{article}
\usepackage{amsmath,amssymb}
\usepackage[dvips]{lscape,graphicx}

\begin{document}

\author{R. B. Nevzorov${}^{\dag,\ddag}$ and M. A. Trusov${}^{\dag}$ \\[5mm] {\itshape ${}^{\dag}$ITEP, Moscow, Russia} \\
{\itshape ${}^{\ddag}$DESY Theory, Hamburg, Germany}}

\title{Renormalization of the soft SUSY breaking terms in the strong Yukawa coupling limit in the NMSSM}

\maketitle

\begin{abstract}
\noindent  In the strong Yukawa coupling limit the renormalization
of trilinear couplings $A_i(t)$ and combinations of the scalar
particle masses $\mathfrak{M}_i^2(t)$ are studied in the framework
of the Next--to--Minimal SUSY Model. The dependence of given
parameters on their initial values at the Grand Unification scale
disappears when the solutions of the renormalization group
equations approach the infrared quasi fixed points. Besides that
at $\tilde{\alpha}_{\text{GUT}}\ll Y_i(0)\ll 1$ in the vicinity of
quasi fixed points $A_i(t)$ and $\mathfrak{M}_i^2(t)$ concentrated
near lines or surfaces in the soft SUSY breaking parameter space.
It occurs because the dependence of trilinear couplings and
combinations of scalar particle mass on $A_i(0)$ and
$\mathfrak{M}_i^2(0)$ disappears rather weakly. We propose a
method which enables to obtain equations for such lines and
surfaces. Their origin is also discussed by means of exact and
approximate solutions of the NMSSM renormalization group
equations.
\end{abstract}

\newpage

\section{Introduction}

Parameters of a soft breaking of supersymmetry plays a key role in
analysing the particle spectra in supersymmetric (SUSY) models.
The Bose--Fermi degeneracy of the spectrum is one of the most
serious flaws in SUSY models. This means that the masses of
observable particles and their superpartners coincide in the limit
of exact supersymmetry, but this is in glaring contradiction with
experimental data. Thus, supersymmetry must be broken, but its
breaking must not lead to the hierarchy problem \cite{1}. Such a
breaking of supersymmetry is referred to as a soft breaking.
Supersymmetry breaking associated with supergravity (SUGRA)
effects is one of the most promising mechanisms for constructing a
realistic model. Although the Lagrangian of $N=1$ SUGRA models
\cite{2} is unrenormalizable, it can be shown that, in the
low--energy region $E\ll M_{\text{Pl}}$, where
$M_{\text{Pl}}=2.4\cdot 10^{18}\text{~GeV}$ is the Planck mass,
all unrenormalizable terms in the Lagrangian for observable fields
are suppressed in a power--law way with respect to
$M_{\text{Pl}}$, vanishing in the limit $M_{\text{Pl}}\to\infty$.
In the case being considered, the Lagrangian for observable fields
can be represented as the sum
\begin{equation}
\mathcal{L}=\mathcal{L}_{\text{SUSY}}+\mathcal{L}_{\text{soft}},
\label{1}
\end{equation}
where $\mathcal{L}_{\text{SUSY}}$ is the Lagrangian corresponding
to unbroken supersymmetry, while $\mathcal{L}_{\text{soft}}$ takes
into account terms that generate a soft breaking of supersymmetry.

To a considerable extent, the form of the first term in (\ref{1})
is determined by the superpotential of the SUSY model under study,
this superpotential in turn being a function of chiral
superfields. For a SUSY theory to be renormalizable, the
superpotential must include only terms that are quadratic and
cubic in chiral superfields $S_\alpha$:
\begin{equation}
W(S_{\alpha})=\frac{1}{2}\mu_{\alpha\beta}S_{\alpha}S_{\beta}+
\frac{1}{6} h_{\alpha\beta\gamma}S_{\alpha}S_{\beta}S_{\gamma}.
\label{2}
\end{equation}
When local supersymmetry is broken in SUGRA models, parameters of
a soft breaking of supersymmetry are generated in the sector of
observable fields \cite{3},\cite{4}. Since the explicit form of
terms that violate supersymmetry, but which do not lead to the
emergence of second--order divergences, is known \cite{5}, the
Lagrangian for the case where supersymmetry is softly broken can
be represented as
\begin{equation}
\mathcal{L}_{\text{soft}}=\frac{1}{2}M_a\bar{\lambda_a}\lambda_a-
m^2_{\alpha}y^*_{\alpha}y_{\alpha}-\left(\frac{1}{6}A_{\alpha\beta\gamma}
h_{\alpha\beta\gamma}y_{\alpha}y_{\beta}y_{\gamma}+\frac{1}{2}B_{\alpha\beta}
\mu_{\alpha\beta}y_{\alpha}y_{\beta}+\text{h.c.}\right),
\label{3}
\end{equation}
where $y_\alpha$ are the scalar components of the chiral
superfields $S_\alpha$ and $\lambda_\alpha$ are gaugino fields.
Usually, the parameters $M_\alpha$, $m_\alpha^2$,
$A_{\alpha\beta\gamma}$, and $B_{\alpha\beta}$ of a soft breaking
of supersymmetry are defined at the scale of $M_X\approx 3\cdot
10^{16}\text{~GeV}$. In minimal SUSY models, the values of all
three gauge coupling constants $g_i$ coincide at this scale:
$g_i(M_X)=g_{\text{GUT}}$ \cite{6}. This relationship between the
gauge coupling constants arises within Grand Unified Theories
\cite{7}. In such theories, all observable gauge bosons and their
superpartners (gaugino fields) belong to the same multiplet;
therefore, the masses of all gauginos also coincide at the scale
$M_X$. In the following, we everywhere set
$M_\alpha(M_X)=M_{1/2}$. The parameters $M_{1/2}$, $m_\alpha^2$,
$A_{\alpha\beta\gamma}$, and $B_{\alpha\beta}$ defined in this way
at the Grand Unification scale should be treated as boundary
conditions for the set of renormalization group equations that
describes the evolution of the parameters of a soft breaking of
supersymmetry down to the electroweak scale.

Within the minimal SUSY Standard Model (MSSM), an exact analytic
solution to the renormalization group equations exists at
$\tan\beta\sim 1$ \cite{8}, in which case the Yukawa coupling
constants $h_b$ and $h_\tau$ for the $b$--quark and
$\tau$--lepton, respectively, are negligibly small. This solution
makes it possible to analyse the evolution of the $t$--quark
Yukawa coupling constant $h_t(t)$ and of the parameters of a soft
breaking of supersymmetry. For the coupling constant $h_t(t)$,
this solution has the form
\begin{equation}
Y_t(t)=\frac{\dfrac{E_t(t)}{6 F_t(t)}}{\left(1+
\dfrac{1}{6Y_t(0)F_t(t)} \right)}, \label{4}
\end{equation}
where $Y_t(t)=h_t^2/(4\pi)^2$ and $t=\ln(M_X^2/Q^2)$. The explicit
expressions for the functions $E_t(t)$ and $F_t(t)$ are presented
in the Appendix (see (\ref{A.2})). At low energies, the second
term in the parentheses on the right--hand side of (\ref{4}) is
much less than unity at sufficiently large values of $h_t^2(0)$;
as a result, all solutions (\ref{4}) to the renormalization group
equations are focused in a narrow interval near the quasi--fixed
point $Y_{\text{QFP}}(t)=E_t(t)/6F_t(t)$ \cite{9}. Formally, a
solution of this type can be obtained by making $Y_t(0)$ tend to
infinity in (\ref{4}).

Along with the Yukawa coupling constant for the $t$--quark,
solution to the renormalization group equations for the
corresponding trilinear coupling constant $A_t$ for scalar fields
and the combination $\mathfrak{M}_t^2=m_Q^2+m_U^2+m_2^2$ of the
scalar--particle masses approach the infrared quasi--fixed point
with increasing $h_t^2(0)$. An analytic solution for these
parameters can be represented as
\begin{equation}
\begin{split}
A_t(t)=&A_t(0)\frac{\epsilon_t(t)}{E_t(t)}+M_{1/2}\left(t
\frac{E_t'(t)}{E_t(t)}-\frac{tE_t(t)-F_t(t)}{F_t(t)}
\left(1-\frac{\epsilon_t(t)}{E_t(t)}\right)\right),\\
\mathfrak{M}_t^2(t)=&\left(\mathfrak{M}_t^2(0)-A_t^2(0)\right)
\frac{\epsilon_t(t)}{E_t(t)}+
\left(A_t(0)\frac{\epsilon_t(t)}{E_t(t)}-M_{1/2}
\frac{tE_t(t)-F_t(t)}{F_t(t)}\right.\\
&\left.{}\times\left(1-\frac{\epsilon_t(t)}{E_t(t)}\right)\right)^2
+M_{1/2}^2\left[\frac{d}{dt}\left(t^2\frac{E'_t(t)}{E_t(t)}
\right)-\frac{t^2 E'_t(t)}{F_t(t)}
\left(1-\frac{\epsilon_t(t)}{E_t(t)}\right)\right],
\end{split}
\label{5}
\end{equation}
where $\epsilon_t(t)=Y_t(t)/Y_t(0)$. In the regime of strong
Yukawa coupling, in which case $h_t^2(0)\gg g_{\text{GUT}}^2(0)$,
the natural small parameter $\epsilon_t(t)$ arises in the theory
being considered. In the infrared region, the dependence of the
solutions in (\ref{4}) and (\ref{5}) on the boundary conditions at
the Grand Unification scale disappears almost completely. Near the
infrared fixed point, we have $Y_t(t_0)\approx
Y_{\text{QFP}}(t_0)$, $A_t(t_0)\approx A_{\text{QFP}}(t_0)$, and
$\mathfrak{M}_t^2(t_0)\approx\mathfrak{M}_{\text{QFP}}^2(t_0)$,
where $A_{\text{QFP}}(t)$ and $\mathfrak{M}_{\text{QFP}}^2(t)$ are
expressed in terms of only the gaugino mass at the scale $M_X$ and
$t_0=2\ln(M_X/M_t^{\text{pole}})$, the pole $t$--quark mass
$M_t^{\text{pole}}$ being approximately equal to $175\text{~GeV}$.
The deviations from $Y_{\text{QFP}}(t)$, $A_{\text{QFP}}(t)$, and
$\mathfrak{M}_{\text{QFP}}^2(t)$ are determined by the ratio
$\epsilon_t(t)/E_t(t)\approx 1/(6F_t(t)Y_t(0))$, which is of order
$1/(10h_t^2(0))$ at the electroweak scale. The properties of
solutions to the renormalization group equations within the MSSM
and the spectrum of particles in the infrared fixed point regime
at $\tan\beta\sim 1$ were investigated in \cite{10}-\cite{13}.

At large values of $\tan\beta$ (about $50$ to $60$), the Yukawa
coupling constants $h_b$ and $h_\tau$ are of order $h_t$; for this
case, an exact analytic solution to the renormalization group
equations has not yet been found. Nonetheless, the detailed
investigations performed in \cite{1},\cite{13},\cite{14} revealed
that, in this region of the parameter space, solution to the set
of nonlinear differential equations being studied also approach
the infrared quasi--fixed point, the basic properties of the
solutions remaining unchanged.

A reduction of the number of independent parameters in the
vicinity of the infrared fixed point at $\tan\beta\sim 1$ make it
possible to obtain, to a sufficiently high degree of precision, an
upper limit on the mass of the lightest Higgs boson. In the case
being considered, this upper limit does not exceed $94\pm
5\text{~GeV}$ \cite{12},\cite{13},\cite{15}, the constraints on
the mass of the lightest Higgs boson from LEP\,II \cite{16} being
such that a considerable part of solutions approaching the
quasi--fixed point at $\tan\beta\sim 1$ have already been ruled
out by the existing experimental data. This gives an additional
incentive to study the Higgs sector and the renormalization group
equations and their solutions within more involved SUSY models.

Of extensions of the MSSM, the simplest one that makes it possible
to preserve the unification of the gauge constants and which leads
to a higher upper limit on the mass of the lightest Higgs boson is
the nonminimal SUSY Standard Model (NMSSM) introduced in
\cite{17},\cite{18}. The Higgs sector of this model contains an
extra singlet superfield $Y$ in addition to two doublets $H_1$ and
$H_2$. The upper limit on the mass of the lightest Higgs boson
within the NMSSM attains a maximum value in the regime of strong
Yukawa coupling, in which case the Yukawa coupling constants
$Y_i(0)$ are much greater than the gauge coupling constant
$\tilde{\alpha}_{\text{GUT}}=g_{\text{GUT}}^2/(4\pi)^2$. In the
parameter space region being considered, solutions to the
renormalization group equations within the NMSSM are attracted to
quasi--fixed lines or surfaces in the space of Yukawa coupling
constants. In the limit $Y_i(0)\to\infty$, all solutions to the
set of differential equations in question are concentrated near
quasi--fixed points \cite{19}, which arise as the result of
intersections of Hill lines or surfaces with the invariant line
connecting the stable fixed point at $Y_i\gg\tilde{\alpha}_i$
\cite{20} with the stable infrared fixed point within the NMSSM
\cite{21}.

For the parameters of a soft breaking of supersymmetry, the
behaviour of solutions to the renormalization group equations
within the NMSSM is studied here near infrared quasi--fixed
points. Approximate solutions for the trilinear coupling constants
$A_i(t)$ and for the combinations $\mathfrak{M}_i^2(t)$ of the
scalar particle masses are presented in the Appendix. In the
regime of strong Yukawa coupling, where
$\tilde{\alpha}_{\text{GUT}}\ll Y_i(0)\ll 1$, $A_i(t)$ and
$\mathfrak{M}_i^2(t)$ at the electroweak scale are concentrated
near some straight lines or surfaces in the space of parameters
being considered. With increasing $Y_i(0)$, the dependence of the
trilinear coupling constants and of the combinations of the scalar
particle masses on $A_i(0)$ and $\mathfrak{M}_i^2(0)$ becomes
weaker; in the limit $Y_i(0)\to\infty$, solutions to the
renormalization group equations for the parameters of a soft
breaking of supersymmetry approach a quasi--fixed point.

\section{Renormalization of the soft SUSY breaking terms for universal boundary conditions}

The superpotential in the NMSSM involves a large number of Yukawa
coupling constants. At $\tan\beta\sim 1$, however, all of these
are small, with the exception of the $t$--quark Yukawa coupling
constant $h_t$, the self--interaction constant $\varkappa$ for the
neutral scalar field $Y$, and the constant $\lambda$
characterising the interaction of the field $Y$ with the doublets
$H_1$ and $H_2$. In the regime of strong Yukawa coupling, the
aforementioned constants can be chosen in such a way that, at the
scale $M_X$, the Yukawa coupling constant $h_b$ for the $b$--quark
is equal to the Yukawa coupling constant $h_\tau$ for the
$\tau$--lepton \cite{19},\cite{22}. This relationship between
$h_b$ and $h_\tau$ is usually realised in the minimal schemes of
unification of gauge interactions \cite{23}.

Disregarding all Yukawa coupling constants, with the exception of
$h_t$, $\lambda$, and $\varkappa$, we can represent the total
superpotential within the NMSSM in the form
\begin{equation}
W=\lambda Y(H_1 H_2)+\frac{\varkappa}{3}Y^3+h_t(H_2Q)U^c_R.
\label{6}
\end{equation}
By construction, the superpotential of the nonminimal SUSY model
is invariant under the discrete transformations $y'_\alpha=e^{2\pi
i/3 }y_\alpha$ of the $Z_3$ group \cite{18}. Second order terms in
superfields do not satisfy this condition; therefore, they have
been eliminated from the superpotential (\ref{6}). Upon a
spontaneous breaking of gauge symmetry, the field $Y$ develops a
nonzero vacuum expectation value ($\langle Y\rangle=y/\sqrt{2}$).
Thus, a mixing of the doublets $H_1$ and $H_2$, which is necessary
for the emergence of the vacuum expectation value $v_1$ of the
$H_1$ doublet (without this vacuum expectation value, down quarks
and charged leptons remain massless), is generated within the
NMSSM.

That the extra superfield $Y$, which is a singlet with respect to
$SU(2)\otimes U(1)$ gauge interactions, and the coupling constant
$\lambda$ are introduced in the superpotential (\ref{6}) leads to
an increase in the upper limit on the mass of the lightest Higgs
boson in relation to that within the MSSM: in the NMSSM, it is
$135\text{~GeV}$ \cite{24}, which is $7-10\text{~GeV}$ greater
than the corresponding value in the minimal SUSY model, the
largest value of the mass of the lightest Higgs boson being
attained in the regime of strong Yukawa coupling.

Upon a soft breaking of supersymmetry due to SUGRA effects, scalar
fields acquire masses $m_i^2$; in addition, a trilinear constant
$A_i$ for the interaction of scalar fields is associated with each
Yukawa coupling constant in the total Lagrangian of the theory.
That the models being considered involve a large number of unknown
parameters of a soft breaking of supersymmetry is one of the main
flaws in such models. The hypothesis of universality of these
constants at the scale $M_X$ makes it possible to reduce their
number in the NMSSM to three (the mass $m_0$ of all scalar
particles, the trilinear coupling constant $A$ for the interaction
of scalar fields, and the gaugino mass $M_{1/2}$), whereby the
analysis of the spectrum of SUSY particles is significantly
simplified. Naturally, the universal parameters of supersymmetry
breaking arise in the minimal SUGRA model (see \cite{4},\cite{25})
and in some string models \cite{26}. Even in the region of low
energies, the hypothesis of universality of fundamental parameters
makes it possible to avoid the emerging of flavour--changing
neutral currents. Thus, the minimal set of the fundamental
parameters in the NMSSM includes, in addition to the constants of
the Standard Model, five unknown constants ($\lambda$,
$\varkappa$, $A$, $m_0$, and $M_{1/2}$). Within the nonminimal
SUSY model, the spectrum of the superpartners of observable
particles and Higgs bosons for universal boundary conditions was
studied in \cite{27}.

The full set of renormalization group equations that describes the
evolution of Yukawa coupling constants, the trilinear coupling
constants $A_i(t)$, and the scalar particle masses $m_i^2(t)$
within the nonminimal SUSY model from the Grand Unification scale
down to the electroweak scale can be found in \cite{28},\cite{29}
(see also Appendix). This set of equations is nonlinear even in
the one--loop approximation; in view of this, its analytic
solution has not yet been found. All equations that form the set
being considered can be partitioned into two groups. The first
includes equations that describe the evolution of gauge and Yukawa
coupling constants. If their evolution is known, the remaining
equations from the set of renormalization group equations can be
considered as a set of linear differential equations for the
parameters of a soft breaking of supersymmetry. In solving this
set of equations, it is necessary to integrate, first of all, the
equations for the gaugino masses and for the trilinear coupling
constants for the interactions of scalar particles. A general
solution to the set of linear differential equations
\[ \frac{dy_i(t)}{dt}=S_{ij}(t)y_{j}(t)+F_i(t), \]
where the matrix $S_{ij}(t)$ and the column vector (nonhomogeneous
term) $F_i(t)$ are known, has the form
\begin{equation}
\label{7}
y_i(t)=\Phi_{ij}(t)y_j(0)+\Phi_{ik}(t)\int\limits_0^t\Phi_{kj}^{-1}(t')F_j(t')dt',
\end{equation}
where $\Phi_{ij}(t)$ is a solution to the homogeneous equation
$d\Phi_{ij}(t)/dt=S_{ik}(t)\Phi_{kj}(t)$ with the boundary
conditions $\Phi_{ij}(0)=\delta_{ij}$. Since we have
$y_i(0)=(A,A,A)$ for $A_i(t)$ if the fundamental parameters are
chosen in a minimal way and since $F_i(t)\sim M_{1/2}$, the
trilinear coupling constants for the interactions of scalar fields
are given by
\begin{equation}
A_i(t)=e_i(t)A+f_i(t)M_{1/2}. \label{8}
\end{equation}
These solutions for $A_i(t)$ must be substituted into the
expressions on the right--hand sides of the renormalization group
equations for the scalar particle masses, the functions $F_i(t)$
involving terms proportional to $A^2$, $AM_{1/2}$, and
$M_{1/2}^2$. Considering that, in the case being considered,
$y_i(0)=(m_0^2,m_0^2,m_0^2,m_0^2,m_0^2)$, we can represent the
required solution for $m_i^2(t)$ as
\begin{equation}
m_i^2(t)=a_i(t)m_0^2+b_i(t)M_{1/2}^2+c_i(t)AM_{1/2}+d_i(t)A^2.
\label{9}
\end{equation}
The functions $e_i(t)$, $f_i(t)$, $a_i(t)$, $b_i(t)$, $c_i(t)$,
and $d_i(t)$, which determine the evolution of $A_i(t)$ and
$m_i^2(t)$, remain unknown, since an analytic solution to the full
set of renormalization group equations within the NMSSM is
unavailable. These functions greatly depend on the choice of
values for the Yukawa coupling constants at the Grand Unification
scale $M_X$. At the electroweak scale $t=t_0$, relations (\ref{8})
and (\ref{9}) specify the parameters $A_i^2(t_0)$ and $m_i^2(t_0)$
of a soft breaking of supersymmetry as functions of their initial
values at the Grand Unification scale.

The results of our numerical analysis, which are presented in the
Table, indicate that, with increasing $Y_i(0)$, where
$Y_t(t)=\dfrac{h_t^2(t)}{(4\pi)^2}$,
$Y_\lambda(t)=\dfrac{\lambda^2(t)}{(4\pi)^2}$, and
$Y_\varkappa(t)=\dfrac{\varkappa^2(t)}{(4\pi)^2}$, the functions
$e_i(t_0)$, $c_i(t_0)$, and $d_i(t_0)$ decrease and tend to zero
in the limit $Y_i(0)\to\infty$, relations (\ref{8}) and (\ref{9})
becoming much simpler in this limit. Instead of the squares of the
scalar particle masses, it is convenient to consider their linear
combinations
\begin{equation}
\begin{split}
\mathfrak{M}_t^2(t)&=m_2^2(t)+m_Q^2(t)+m_U^2(t),\\
\mathfrak{M}_{\lambda}^2(t)&=m_1^2(t)+m_2^2(t)+m_y^2(t),\\
\mathfrak{M}_{\varkappa}^2(t)&=3m_y^2(t)
\end{split}
\label{10}
\end{equation}
in analysing the set of renormalization group equations. In the
case of universal boundary conditions, the solutions to the
differential equations for $\mathfrak{M}_i^2(t)$ can be
represented in the same form as the solutions for $m_i^2(t)$ (see
(\ref{9})); that is
\begin{equation}
\mathfrak{M}_i^2(t)=3\tilde{a}_i(t)m_0^2+\tilde{b}_i(t)M_{1/2}^2+
\tilde{c}_i(t)A M_{1/2}+\tilde{d}_i(t)A^2. \label{11}
\end{equation}
Since the homogeneous equations for $A_i(t)$ and
$\mathfrak{M}_i^2(t)$ have the same form, the functions
$\tilde{a}_i(t)$ and $e_i(t)$ coincide; in the limit of strong
Yukawa coupling, the $m_0^2$ dependence disappears in the
combinations (\ref{10}) of the scalar particle masses as the
solutions to the renormalization group equations for the Yukawa
coupling constants approach quasi--fixed points. This behaviour of
the solutions implies that $A_i(t)$ and $\mathfrak{M}_i^2(t)$
corresponding to $Y_i(0)\gg\tilde{\alpha}_i(0)$ also approach
quasi--fixed points. As was shown in \cite{19}, two quasi--fixed
points of the renormalization group equations within the NMSSM are
of greatest interest from the physical point of view. Of these,
one corresponds to the boundary conditions
$Y_t(0)=Y_\lambda(0)\gg\tilde{\alpha}_i(0)$ and $Y_\varkappa(0)=0$
for the Yukawa coupling constants. The fixed points calculated for
the parameters of a soft breaking of supersymmetry by using these
values of the Yukawa coupling constants are
\begin{equation}
\begin{aligned} \rho_{A_t}^{\text{QFP}}(t_0)&\approx 1.77,
&\rho_{\mathfrak{M}^2_t}^{\text{QFP}}(t_0)&\approx 6.09,\\
\rho_{A_{\lambda}}^{\text{QFP}}(t_0)&\approx -0.42,\qquad
&\rho_{\mathfrak{M}^2_{\lambda}}^{\text{QFP}}(t_0)&\approx -2.28,
\end{aligned}
\label{12}
\end{equation}
where $\rho_{A_i}(t)=A_i(t)/M_{1/2}$ and
$\rho_{\mathfrak{M}_i^2}(t)=\mathfrak{M}_i^2/M_{1/2}^2$. Since the
coupling constant $\varkappa$ for the self--interaction of neutral
scalar fields is small in the case being considered,
$A_\varkappa(t)$ and $\mathfrak{M}_\varkappa^2(t)$ do not approach
the quasi--fixed point. Nonetheless, the spectrum of SUSY
particles is virtually independent of the trilinear coupling
constant $A_\varkappa$ since $\varkappa\to 0$.

In just the same way, one can determine the position of the other
quasi--fixed point for $A_i(t)$ and $\mathfrak{M}_i^2(t)$, that
which corresponds to $R_{\lambda 0}=3/4,~R_{\varkappa 0}=3/8$. The
results are
\begin{equation}
\begin{aligned} \rho_{A_t}^{\text{QFP}}(t_0)&\approx 1.73,
&\rho_{A_{\lambda}}^{\text{QFP}}(t_0)&\approx -0.43,
&\rho_{A_{\varkappa}}^{\text{QFP}}(t_0)&\approx 0.033,\\
\rho_{\mathfrak{M}^2_t}^{\text{QFP}}(t_0)&\approx 6.02,\quad
&\rho_{\mathfrak{M}^2_{\lambda}}^{\text{QFP}}(t_0)&\approx
-2.34,\quad
&\rho_{\mathfrak{M}^2_{\varkappa}}^{\text{QFP}}(t_0)&\approx 0.29,
\end{aligned}
\label{13}
\end{equation}
where $R_{\lambda 0}=Y_\lambda(0)/Y_t(0)$ and $R_{\varkappa
0}=Y_\varkappa(0)/Y_t(0)$. It should be noted that, in the
vicinities of quasi--fixed points, we have
$\rho_{\mathfrak{M}^2_{\lambda}}^{\text{QFP}}(t_0)<0$. Negative
values of $\mathfrak{M}_\lambda^2(t_0)$ lead to a negative value
of the parameter $m_2^2(t_0)$ in the potential of interaction of
Higgs fields (see Table). In other words, an elegant mechanism
that is responsible for a radiative violation of $SU(2)\otimes
U(1)$ symmetry and which does not require introducing tachyons in
the spectrum of the theory from the outset survives in the regime
of strong Yukawa coupling within the NMSSM. This mechanism of
gauge symmetry breaking was first discussed in \cite{30} by
considering the example of the minimal SUSY model.

The evolution of the constants of a soft breaking of
supersymmetry, $A_i(t)$ and $\mathfrak{M}_i^2(t)$, is illustrated
in Figs. 1-4 for $Y_t(0),Y_\lambda(0)\gg\tilde{\alpha}(0)$ and
$Y_\varkappa=0$, as well as for
$Y_\varkappa(0)\gg\tilde{\alpha}(0)$. Although the ratio
$A/M_{1/2}$ is varied between $-1$ and $1$, with the scalar
particle mass $m_0^2$ lying in the range $0\le m_0^2\le
M_{1/2}^2$, solutions to the renormalization group equations are
focused in a narrow interval at low energies. However,
$A_\varkappa(t)$ and $\mathfrak{M}_\varkappa^2(t)$ are
concentrated near zero, since the neutral field $Y$ is not
renormalized by gauge interactions; therefore, they remain
dependent on the initial conditions (see also Table). The values
of $A_t(t_0)$ and $\mathfrak{M}_t^2(t_0)$ show the weakest
dependence on $A$ and $m_0^2$, because the former are renormalized
by strong interactions.

By using the fact that $\mathfrak{M}_i^2(t)$ as determined for the
case of universal boundary conditions is virtually independent of
$m_0^2$, we can predict $a_i(t_0)$ values near the quasi--fixed
points. The results are
\begin{equation}
\begin{split}
1)~&R_{\lambda 0}=1,~R_{\varkappa 0}=0,\\
&a_y(t_0)=a_u(t_0)=\frac{1}{7},~a_1(t_0)=a_q(t_0)=\frac{4}{7},~
a_2(t_0)=-\frac{5}{7}\, ;\\ 2)~&R_{\lambda 0}=3/4,~R_{\varkappa
0}=3/8,\\ &a_y(t_0)=0,~
a_1(t_0)=-a_2(t_0)=\frac{2}{3},~a_q(t_0)=\frac{5}{9},~
a_u(t_0)=\frac{1}{9}\, .
\end{split}
\label{14}
\end{equation}
To do this, it was necessary to consider specific combinations of
the scalar particle masses, such as $m_U^2-2m_Q^2$,
$m_Q^2+m_U^2-m_2^2+m_1^2$, and $m_y^2-2m_1^2$ (at $\varkappa=0$),
that are not renormalized by Yukawa interactions. As a result, the
dependence of the above combinations of the scalar particle masses
on $m_0^2$ at the electroweak scale is identical to that at the
Grand Unification scale. The data in the Table show that the
predictions in (\ref{14}) agree fairly well with the results of
numerical calculations.

\section{Lines and planes in the space of parameters of the soft SUSY breaking}

By considering the example of an exact analytic solution
$Y_\lambda=0$, we can see how the $A$ and $m_0^2$ dependence of
the parameters of the parameters of a soft breaking of
supersymmetry disappears. From this point, our analysis of
solutions to renormalization group equations for $A_i(t)$ and
$\mathfrak{M}_i^2(t)$ will not be restricted to the case of
universal boundary conditions; that is, $A_i(0)$ and
$\mathfrak{M}_i^2(0)$ will be considered as independent boundary
values. In the case being considered, the full set of
renormalization group equations within the NMSSM can be
partitioned into two subsets -- of these, one coincides with the
set of renormalization group equations within the MSSM, while the
other describes the evolution of $Y_\varkappa(t)$,
$A_\varkappa(t)$, and $\mathfrak{M}_\varkappa^2(t)$. For
$Y_\lambda=0$, the evolution of $A_t(t)$ and $\mathfrak{M}_t^2(t)$
is determined by relations (\ref{5}). For the other two parameters
$A_\varkappa(t)$ and $\mathfrak{M}_\varkappa^2(t)$ of a soft
breaking of supersymmetry, which correspond to the Yukawa coupling
constant $\varkappa$, we obtain
\begin{equation}
\begin{split}
A_{\varkappa}(t)&=A_{\varkappa}(0)\epsilon_{\varkappa}(t),\\
\mathfrak{M}^2_{\varkappa}(t)&=\mathfrak{M}^2_{\varkappa}(0)\epsilon_{\varkappa}(t)-
A^2_{\varkappa}(0)
\epsilon_{\varkappa}(t)(1-\epsilon_{\varkappa}(t)),
\end{split}
\label{15}
\end{equation}
where
$\epsilon_\varkappa(t)=Y_\varkappa(t)/Y_\varkappa(0)=1/(1+6Y_\varkappa(0)t)$.
With increasing $Y_\varkappa(0)$ and $Y_t(0)$, the values of the
Yukawa coupling constants at the electroweak scale approach the
quasi--fixed point, while solutions to the renormalization group
equations for the parameters of a soft breaking of supersymmetry
are attracted to the straight lines $A_t=1.67M_{1/2}$ and
$A_\varkappa=0$ in the $(A_t,A_\varkappa)$ plane and to the
straight lines $\mathfrak{M}_t^2=5.49M_{1/2}^2$ and
$\mathfrak{M}_\varkappa^2=0$ in the
$(\mathfrak{M}_t^2,\mathfrak{M}_\varkappa^2)$ plane. The
dependence on the boundary values $A_i(0)$ and
$\mathfrak{M}_i^2(0)$ disappears only in the limit
$Y_i(0)\to\infty$, which corresponds to a quasi--fixed point in
the $(\rho_t,\rho_\varkappa)$ plane.

At nonzero values of $Y_\lambda$, it is possible to construct, on
the basis of an approximate solution for the Yukawa coupling
constants within the NMSSM \cite{19}, approximate solutions to the
renormalization group equations for the parameters of a soft
breaking of supersymmetry. Near the quasi--fixed points given by
(\ref{12}) and $\ref{13}$, these solutions describe the evolution
of $A_t(t)$ and $\mathfrak{M}_t^2(t)$ to a fairly high precision
(about 1\%). The relative deviations of the approximate solutions
from numerical ones are significantly greater for $A_\lambda(t)$
and $\mathfrak{M}_\lambda^2(t)$. At the electroweak scale, the
relative error here is as large as $20-30\%$ near quasi--fixed
points. Finally, the approximate solutions for the parameters of a
soft breaking of supersymmetry that correspond to the Yukawa
coupling constant $\varkappa$ are characterised by the poorest
accuracy, providing only a qualitative description of the
behaviour of the numerical solutions. Our results lead to the
conclusion that the relative deviation of the approximate solution
being studied from the precise numerical solution decreases with
increasing contribution of gauge interactions to the
renormalization of the corresponding parameter of a soft breaking
of supersymmetry.

Approximate solutions to the renormalization group equations
within the NMSSM for the trilinear coupling constants and the
combination (\ref{10}) of the scalar particle masses are presented
in the Appendix. In the regime of strong Yukawa coupling at
$A_i(0)=A$ and $\mathfrak{M}_i^2(0)=3m_0^2$, the dependence of
these solutions on the initial conditions vanishes in proportion
to $1/Y_i(0)$.  Disregarding $O(1/Y_i(0))$ terms, we find, for
nonuniversal boundary conditions, that
\begin{equation}
\begin{split}
\binom{A_t(t)}{\mathfrak{M}^2_t(t)}\approx & \frac{R_{\lambda 0}
F_{\lambda}(t)}{6F_t(t)+2R_{\lambda 0}F_{\lambda}(t)}
\binom{A_t(0)-A_{\lambda}(0)}{\mathfrak{M}^2_t(0)-\mathfrak{M}^2_{\lambda}(0)}+\dots
,\\ \binom{A_{\lambda}(t)}{\mathfrak{M}^2_{\lambda}(t)}\approx &
\frac{3 F_t(t)}{6F_t(t)+2R_{\lambda 0}F_{\lambda}(t)}
\binom{A_{\lambda}(0)-A_t(0)}
{\mathfrak{M}^2_{\lambda}(0)-\mathfrak{M}^2_t(0)} \\
&{}+\frac{2R_{\varkappa 0}t}{6R_{\lambda 0}F_{\lambda}(t)+
6R_{\varkappa 0}t} \binom{A_{\lambda}(0)-A_{\varkappa}(0)}
{\mathfrak{M}^2_{\lambda}(0)- \mathfrak{M}^2_{\varkappa}(0)}+\dots
,\\ \binom{A_{\varkappa}(t)}{\mathfrak{M}^2_{\varkappa}(t)}\approx
&  \frac{6 R_{\lambda 0} F_{\lambda}(t)}{6R_{\lambda
0}F_{\lambda}(t)+6R_{\varkappa 0}t}
\binom{A_{\varkappa}(0)-A_{\lambda}(0)}
{\mathfrak{M}^2_{\varkappa}(0)- \mathfrak{M}^2_{\lambda}(0)}+\dots
.
\end{split}
\label{16}
\end{equation}
In (\ref{16}), we discarded terms proportional to $M_{1/2}$,
$M_{1/2}^2$, $A_i(0)M_{1/2}$, and $A_i(0)A_j(0)$. The explicit
expressions for the functions $F_i(t)$ are presented in the
Appendix (see (\ref{A.2})). From the results that we obtained, it
follows that, in the case of nonuniversal boundary conditions, the
approximate solutions to the renormalization group equations for
$A_i(t)$ and $\mathfrak{M}_i^2(t)$ remain dependent on the
difference of the values of these constants at the Grand
Unification scale. At the same time, the linear combinations
\begin{equation}
\begin{gathered}
9F_t(t)A_t(t)+3R_{\lambda 0}F_{\lambda}(t)A_{\lambda}(t)+
R_{\varkappa 0}tA_{\varkappa}(t)=const,\\
9F_t(t)\mathfrak{M}^2_t(t)+3R_{\lambda 0}F_{\lambda}(t)
\mathfrak{M}^2_{\lambda}(t)+R_{\varkappa
0}t\mathfrak{M}^2_{\varkappa}(t)=const',
\end{gathered}
\label{17}
\end{equation}
of the parameters undergo virtually no changes in response to
variations of $A_i(0)$ in the first combination and of
$\mathfrak{M}_i^2(0)$ in the second combination. At the
electroweak scale ($t=t_0$), relations (\ref{17}) specify straight
lines (at $Y_\varkappa=0$) and planes (at
$Y_\varkappa(0)\gg\tilde{\alpha}(0)$) in the space of the
parameters of a soft breaking of supersymmetry. In the regime of
strong Yukawa coupling, the approximate solutions to the
renormalization group equations are attracted to these straight
lines and planes.

The results of our numerical analysis, which are illustrated in
Figs. 5 and 6, indicate that, in the vicinity of the infrared
fixed point at $Y_\varkappa=0$, solutions to the renormalization
group equations at the electroweak scale are indeed concentrated
near some straight lines for the case where the simulation was
performed by using boundary conditions uniformly distributed in
the $(A_t,A_\lambda)$ and the
$(\mathfrak{M}_t^2,\mathfrak{M}_\lambda^2)$ plane. The strength
with which these solutions are attracted to them grows with
increasing $Y_i(0)$. The equations for the lines being considered
can be obtained by fitting the numerical results displayed in
Figs. 5 and 6. This yields
\begin{equation}
\begin{gathered}
A_t+0.147(0.121)A_{\lambda}=1.70M_{1/2},\\
\mathfrak{M}^2_t+0.147(0.121)\mathfrak{M}^2_{\lambda}=5.76
M_{1/2}^2 .
\end{gathered}
\label{18}
\end{equation}
For $Y_\varkappa(0)\gg\tilde{\alpha}_i(0)$ solutions to the
renormalization group equations are grouped near planes in the
space of the parameters of a soft breaking of supersymmetry
$(A_t,A_\lambda,A_\varkappa)$ and
$(\mathfrak{M}_t^2,\mathfrak{M}_\lambda^2,\mathfrak{M}_\varkappa^2)$
(see Figs. 7-9):
\begin{equation}
\begin{gathered}
A_t+0.128(0.091)A_{\lambda}+0.022(0.0105)A_{\varkappa}=1.68
M_{1/2} ,\\
\mathfrak{M}^2_t+0.128(0.091)\mathfrak{M}^2_{\lambda}+0.022(0.0105)
\mathfrak{M}^2_{\varkappa}=5.77 M_{1/2}^2 .
\end{gathered}
\label{19}
\end{equation}
In (\ref{18}) and (\ref{19}), the predictions for the coefficients
of $A_\lambda$, $A_\varkappa$, $\mathfrak{M}_\lambda^2$, and
$\mathfrak{M}_\varkappa^2$ according to the calculations at
$R_{\lambda 0}=1,~R_{\varkappa 0}=0$ and at $R_{\lambda
0}=3/4,~R_{\varkappa 0}=3/8$ on the basis of the approximate
solutions (\ref{16}) and (\ref{17}) to the renormalization group
equations within the NMSSM in the regime of strong Yukawa coupling
are given in parentheses. It can be seen from Figs. 7 and 8 that,
as the values of the Yukawa coupling constants at the Grand
Unification scale are increased, the areas of the surfaces near
which the solutions $A_i(t)$ and $\mathfrak{M}_i^2(t)$ are
concentrated shrink in one of the directions, with the result
that, at $Y_i(0)\sim 1$, the solutions to the renormalization
group equations are attracted to one of the straight lines
belonging to these surfaces.

Thus, the approximate solutions presented in this study lead to
qualitatively correct results. However, an analysis of numerical
solutions to the renormalization group equations revealed that,
with increasing $Y_i(0)$, only in the regime of infrared
quasi--fixed points (that is, at $R_{\lambda 0}=1,~R_{\varkappa
0}=0$ or at $R_{\lambda 0}=3/4,~R_{\varkappa 0}=3/8$) do
$e_i(t_0)$ and $\tilde{a}_i(t_0)$ decrease in proportion to
$1/Y_i(0)$. Otherwise, the dependence on $A$ and $m_0^2$
disappears much more slowly with increasing values of the Yukawa
coupling constants at the Grand Unification scale -- specifically,
in proportion to $(Y_i(0))^{-\delta}$, where $\delta<1$ (for
example, $\delta=0.35-0.40$ at $\varkappa=0$). In the case of
nonuniversal boundary conditions, only when solutions to the
renormalization group equations approach quasi--fixed points are
these solutions attracted to the fixed lines and surfaces in the
space of the parameters of a soft breaking of supersymmetry; in
the limit $Y_i(0)\to\infty$, the parameters $A_i(t)$ and
$\mathfrak{M}_i ^2(t)$ cease to be dependent on the boundary
conditions, in contradiction with the results presented in 16
(\ref{16}). Finally, we note that, by using approximate solutions
to the renormalization group equations within the NMSSM, it is
impossible to explain the emergence of those lines in the
$(A_t,A_\lambda,A_\varkappa)$ and
$(\mathfrak{M}_t^2,\mathfrak{M}_\lambda^2,\mathfrak{M}_\varkappa^2)$
spaces to which solutions of the equations in question are
attracted when all three Yukawa coupling constants are $Y_i(0)\sim
1$.

\section{Analysis of solutions of the renormalization group equations near quasi--fixed points}

That, in the regime of strong Yukawa coupling, quasi--fixed
points, lines, and surfaces appear in the space spanned by the
parameters of a soft a breaking of supersymmetry is explained by
fast changes in the Yukawa coupling constants at the initial
stage. In view of this, the entire scale from $M_X$ to
$M_t^{\text{pole}}$ can be partitioned into two unequal intervals.
Within the initial stage corresponding to $M_X\ge q\ge q_1\sim
10^{12}\text{~GeV}$, the Yukawa coupling constants are
significantly greater than the gauge coupling constants and
decrease fast with increasing $t$. In this region, the evolution
of all fundamental parameters of a soft breaking of supersymmetry
is determined by $Y_i(t)$, so that all gauge coupling constants
$\tilde{\alpha}_i(t)$ can be disregarded for a first
approximation. At the initial stage of evolution, the one--loop
renormalization group equations for $Y_i(t)$ can then be
represented in the form
\begin{equation}
\frac{dY_i(t)}{dt}=-Y_i(t)\left(\sum_j c_{ij}Y_j(t)\right),
\label{20}
\end{equation}
where $c_{ij}\ge 0$. If the set of (\ref{20}) has a fixed point,
all the Yukawa coupling constants are proportional to each other
in the vicinity of this point. In analysing (\ref{20}), it is
therefore convenient to introduce the additional constant $Y_0$
satisfying the equation
\[ \frac{dY_0(t)}{dt}=-Y^2_0(t), \]
and to investigate, instead of $Y_i(t)$, the ratio
$r_i(t)=Y_i(t)/Y_0(t)$. The fixed point of the renormalization
group equations (\ref{20}) is then determined by solving the set
of linear algebraic equations
\begin{equation}
\sum_j\left(c_{ij}r^0_j-\delta_{ij}\right)=0. \label{21}
\end{equation}

Linearising the set of renormalization group (\ref{20}) near the
fixed point $r_i(t)=r_i^0(1+\theta_i(t))$, we obtain
\begin{equation}
\begin{split}
\frac{d\theta_i}{dz}=&\sum_j c_{ij}r^0_j\theta_j,\\
\frac{dA_i}{dz}=&\sum_j\left(c_{ij}r^0_jA_j+
c_{ij}r^0_j\theta_jA_j\right),\\
\frac{d\mathfrak{M}^2_i}{dz}=&\sum_j\left(c_{ij}r^0_j
\left(\mathfrak{M}^2_j+A^2_j\right)+
c_{ij}r^0_j\theta_j\left(\mathfrak{M}^2_j+A^2_j\right)\right),
\end{split}
\label{22}
\end{equation}
where $z=\ln(Y_0(t)/Y_0(0))=-\ln(1+Y_0t)$. A general solution to
this set of linear differential equations at $\theta_i=0$ has the
form
\begin{equation}
\begin{split}
A_i(z)&=\sum_k\alpha_k u_{ik}e^{\lambda_k z},\\
\mathfrak{M}^2_i(z)&=\sum_k\beta_k u_{ik}e^{\lambda_k z}+
\sum_{k,n}\lambda_k e^{\lambda_k
z}u_{ik}u^{-1}_{kn}\int\limits_0^z A^2_n(y) e^{-\lambda_n y}dy,
\end{split}
\label{23}
\end{equation}
where $\lambda_k$ are the eigenvalues of the matrix $c_{ij}r_j^0$,
$u_{ik}$ is the eigenvector corresponding to the eigenvalue
$\lambda_k$, and $\alpha_k$ and $\beta_k$ are constants of
integration that are expressed in terms of $A_i(0)$ and
$\mathfrak{M}_i^2(0)$. It should be noted that all eigenvalues
$\lambda_k$ are positive. They are determined by solving the
characteristic equation $\det(c_{ij}r_j^0-\lambda'\delta_{ij})=0$.

From (\ref{21}), it follows that, at $\lambda'=1$, one of the
columns of the matrix $(c_{ij}r_j^0-\lambda'\delta_{ij})$ is a
linear combination of all other columns; hence, its determinant is
identically equal to zero. This means, that the value of
$\lambda_0=1$ is a root of the characteristic equation. The
eigenvector $u_{i0}=(1,\dots,1)$ corresponds to this eigenvalue.
As a rule, all other roots of the characteristic equation are less
than unity. With increasing $t$, the dependence of solutions
(\ref{23}) on the boundary conditions at the Grand Unification
scale becomes weaker. At $q\sim 10^{12}-10^{13}\text{~GeV}$, in
which case the Yukawa coupling constants are on the same order of
magnitude as the gauge coupling constants, the values of $A_i(z)$
and $\mathfrak{M}_i^2(z)$ at this scale that are calculated by
formulas (\ref{23}) should be treated as new boundary values for
the full set of renormalization group equations within the NMSSM,
which is presented in the Appendix.

In the case where only the $t$--quark Yukawa coupling constant is
nonzero, the set of differential (\ref{20}) for the Yukawa
coupling constants reduces to one equation; as a result, the full
nonlinearised set of renormalization group equations can be solved
exactly in the limit $\tilde{\alpha}_i\to 0$. The results are
given by
\begin{equation}
\begin{gathered}
Y_t(t)=\frac{Y_t(0)}{1+6Y_t(0)t},\quad
A_t(t)=A_t(0)\epsilon_t(t),\\
\mathfrak{M}_t^2(t)=\mathfrak{M}^2_t(0)\epsilon_t(t)-A_t^2(0)\epsilon_t(t)
(1-\epsilon_t(t)),
\end{gathered}
\label{24}
\end{equation}
where $\epsilon_t(t)=Y_t(t)/Y_t(0)$. The solution in (\ref{24})
coincides in form with solution (\ref{23}) to the above linearised
set of differential equations. Formally, it can be obtained by
setting $M_{1/2}=0$ and $E(t)=1$ in the exact solution to the
NMSSM renormalization group equations that is given by (\ref{4})
and (\ref{5}). A comparison of the analytic solution (\ref{24}) in
the regime of strong Yukawa coupling with the exact solution to
the renormalization group equations within the minimal SUSY model
reveals that, as the result of the evolution of $A_t(t)$ and
$\mathfrak{M}_t^2(t)$ from the scale $q_1\sim 10^{12}\text{~GeV}$
to the electroweak scale, there arises, in addition to
$\epsilon_t(t)$, an extra factor $1/E(t)$ in front of $A_t(0)$ and
$\mathfrak{M}_t^2(0)$ in (\ref{5}). This new factor is independent
of the initial conditions for the $t$--quark Yukawa coupling
constant at the Grand Unification scale.

From relations (\ref{23}), it follows that an expansion of
solutions to the renormalization group equations in powers of the
small parameters $\epsilon_t(t)$ -- this expansion emerges in the
regime of strong Yukawa coupling within the MSSM -- is also valid
in more involved models. At the electroweak scale, solution
(\ref{23}) to the renormalization group equations can be
represented as
\begin{equation}
\binom{A_i(t)}{\mathfrak{M}^2_i(t)}=\sum_k u_{ik}v_{ik}(t)
\binom{\alpha_k}{\beta_k}(\epsilon_t(t))^{\lambda_k}+\dots,
\label{25}
\end{equation}
where the quantities $v_{ik}(t)$ are weakly dependent on the
Yukawa coupling constants at the Grand Unification scale and
$\epsilon_t(t)=Y_t(t)/Y_t(0)$. In expressions (\ref{25}), we have
discarded terms proportional to $M_{1/2}$, $M_{1/2}^2$,
$A_iM_{1/2}$, and $A_i^2$. The extra factors $v_{ik}(t)$ appear
upon renormalizing the parameters of a soft breaking of
supersymmetry from $q\sim 10^{12}-10^{13}\text{~GeV}$ to $q\sim
m_t$. At $\varkappa=0$, there exists, in addition to $\lambda_0=1$
and $u_{i0}=(1,1)$, yet another eigenvalue $\lambda_1=3/7$ and the
corresponding eigenvector $u_{i1}=(1,-3)$, whose components
specify $(A_t,A_\lambda)$ and
$(\mathfrak{M}_t^2,\mathfrak{M}_\lambda^2)$. With increasing
$Y_t(0)=Y_\lambda(0)$, the dependence on $\alpha_0$ and $\beta_0$
becomes weaker and the solutions at $t=t_0$ are concentrated near
the straight lines $(A_t(\alpha_1),A_\lambda(\alpha_1))$ and
$(\mathfrak{M}_t^2(\beta_1),\mathfrak{M}_\lambda^2(\beta_1))$. In
order to obtain the equations for these straight lines, it is
necessary to set $A_\lambda(0)=-3A_t(0)$ and
$\mathfrak{M}_\lambda^2(0)=-3\mathfrak{M}_t^2(0)$ at the Grand
Unification scale. At the electroweak scale, there then arise a
relation between $A_t(t_0)$ and $A_\lambda(t_0)$ and a relation
between $\mathfrak{M}_t^2(t_0)$ and $\mathfrak{M}_\lambda^2(t_0)$:
\begin{equation}
\begin{gathered}
A_t+0.137 A_{\lambda}=1.70 M_{1/2},\\ \mathfrak{M}^2_t+ 0.137
\mathfrak{M}^2_{\lambda}=5.76 M^2_{1/2},
\end{gathered}
\label{26}
\end{equation}
These relations agree well with the results obtained above and
presented in (\ref{18}). Moreover, the equations deduced for the
straight lines at $Y_i(0)\sim 1$ by fitting the results of the
numerical of the numerical calculations nearly coincide with those
in (\ref{26}).

When the Yukawa coupling constant $\varkappa$ is nonzero, the
matrix $c_{ij}r_j^0$ has three eigenvalues. One of these is
$\lambda_0=1$, while the other two are
$\lambda_1=\dfrac{3+\sqrt{5}}{9}\approx 0.58$ and
$\lambda_2=\dfrac{3-\sqrt{5}}{9}\approx 0.085$. The corresponding
eigenvectors $u_{i1}$ and $u_{i2}$, which specify
$(A_t,A_\lambda,A_\varkappa)$ and
$(\mathfrak{M}_t^2,\mathfrak{M}_\lambda^2,\mathfrak{M}_\varkappa^2)$
are given by
\begin{equation}
u_{i1}=\begin{pmatrix} -\frac{1}{24}(1+\sqrt{5}) \\
\frac{\sqrt{5}}{6} \\ 1
\end{pmatrix}, \qquad u_{i2}=\begin{pmatrix} \frac{1}{24}(\sqrt{5}-1) \\
-\frac{\sqrt{5}}{6} \\ 1
\end{pmatrix}.
\label{27}
\end{equation}
An increase in $Y_\lambda(0)=2Y_\varkappa(0)=\dfrac{3}{4}Y_t(0)$
leads to the following: first, the dependence of $A_i(t)$ and
$\mathfrak{M}_i^2(t)$ on $\alpha_0$ and $\beta_0$ disappears,
which leads to the emergence of planes in the space spanned by the
parameters of a soft breaking of supersymmetry,
\begin{equation}
\begin{gathered}
A_t+0.103 A_{\lambda}+0.0124 A_{\varkappa}=1.69 M_{1/2},\\
\mathfrak{M}^2_t+0.103 \mathfrak{M}^2_{\lambda}+0.0124
\mathfrak{M}^2_{\varkappa}=5.78 M^2_{1/2};
\end{gathered}
\label{28}
\end{equation}
after that, the dependence on $\alpha_1$ and $\beta_1$ becomes
weaker at $Y_i(0)\sim 1$. This means that, with increasing initial
values of the Yukawa coupling constants, solutions to the
renormalization group equations are grouped near some straight
lines and we can indeed see precisely this pattern in Figs. 7-9.
All equations presented here for the straight lines and planes in
the $\mathfrak{M}_i^2$ space were obtained at $A_i(0)=0$.

From relations (\ref{26}) and (\ref{28}), it follows that
$A_t(t_0)$ and $\mathfrak{M}_t^2(t_0)$ are virtually independent
of the initial conditions; that is, the straight lines and planes
and orthogonal to the $A_t$ and $\mathfrak{M}_t^2$ axes. On the
other hand, the $A_\varkappa(t_0)$ and
$\mathfrak{M}_\varkappa^2(t_0)$ values that correspond to the
Yukawa self--interaction constant $Y$ for the neutral fields are
fully determined by the boundary conditions for the parameters of
a soft breaking of supersymmetry. For this reason, the planes in
the $(A_t,A_\lambda,A_\varkappa)$ and
$(\mathfrak{M}_t^2,\mathfrak{M}_\lambda^2,\mathfrak{M}_\varkappa^2)$
spaces are virtually parallel to the $A_\varkappa$ and
$\mathfrak{M}_\varkappa^2$ axes.

In the case of nonuniversal boundary conditions, the parameters of
a soft breaking of supersymmetry as functions of $A_i(0)$ and
$\mathfrak{M}_i^2(0)$ in he regime of strong Yukawa coupling are
determined by $(\epsilon_t(t))^{\lambda_{\text{min}}}$ (see
(\ref{25})), where $\lambda_{\text{min}}$ is the smallest
eigenvalue of the matrix $c_{ij}r_j^0$ and $\epsilon_t(t_0)\sim
1/h_t^2(0)$. If the values of all $A_i(0)$ and all
$\mathfrak{M}_i^2(0)$ coincide, a solution to be linearised set of
differential equations (\ref{22}) at $\theta_i\ne 0$ can be
represented in the form
\begin{equation}
\begin{split}
\theta_i(z)=&\sum_k\gamma_k u_{ik}e^{\lambda_k z},\\
A_i(z)=&A(0)\left(e^z+\left(e^z-1\right)
\frac{d\theta_i(z)}{dz}\right),\\
\mathfrak{M}^2_i(z)=&\mathfrak{M}^2(0)\left(e^z+\left(e^z-
1\right) \frac{d\theta_i(z)}{dz}\right)+A^2(0)\left(e^z-1\right)
\\
&\times\left(e^z+\frac{1}{2}\left(e^z+1\right)\frac{d\theta_i(z)}{dz}
+\left(e^z-1\right)\frac{d^2\theta_i(z)}{dz^2} \right),
\end{split}
\label{29}
\end{equation}
where $\gamma_k$ are constants of integration. In the stable fixed
point regime, where all $\theta_i$ are equal to zero ,the
evolution of $A_i(t)$ and $\mathfrak{M}_i^2(t)$ is identical to
that within the MSSM (see (\ref{24})). In the limit being
considered, the dependence on $A(0)$ and $\mathfrak{M}^2(0)$ dies
out fast, in proportion to $Y_t(t)/Y_t(0)$, and the parameters
$A_i(t)$ and $\mathfrak{M}_i^2(t)$ approach the fixed point. Since
$\theta_i\sim e^{\lambda_{\text{min}}z}$ for $z\to -\infty$, the
dependence of the parameters of a soft breaking of supersymmetry
on the initial conditions is determined by
$(\epsilon_t(t))^{\lambda_{\text{min}}}$ in all other cases --
that is, by the strength with which the quasi--fixed point
attracts the solutions to the renormalization group equations for
the Yukawa coupling constants in the limiting case of
$Y_i(0)\gg\tilde{\alpha}_i(0)$. As a rule, $\lambda_{\text{min}}$
is significantly less than unity; for example,
$\lambda_{\text{min}}=3/7\approx 0.43$ at $\varkappa=0$ and
$\lambda_{\text{min}}\approx 0.085$ at nonzero $\varkappa$. This
is the reason why, in the regime of strong Yukawa coupling, the
dependence on $A$ and $m_0^2$ dies out at a rate much lower than
that which is predicted by approximate solutions. Nonetheless,
expressions (\ref{23})-(\ref{25}) and (\ref{29}) for $A_i(t)$ and
$\mathfrak{M}_i^2(t)$ tend to zero in the limit $Y_i(0)\to\infty$;
for this reason, the $A_i(0)$ and $\mathfrak{M}_i^2(0)$ dependence
of the solutions being studied disappears completely.

\section{Conclusion}

An increase in the Yukawa coupling constants at the Grand
Unification scale results in that, for $Y_t(t)$, $Y_\lambda(t)$,
and $Y_\varkappa(t)$, the solutions to the renormalization group
equations within the NMSSM are attracted to infrared quasi--fixed
points. We have shown that, as $Y_i(t)$ approach these points, the
corresponding solutions for the trilinear coupling constants
$A_i(t)$ characterising scalar fields and for the combinations
$\mathfrak{M}_i^2(t)$ of the scalar particle masses (see
(\ref{10})) cease to be dependent on their initial values at the
scale $M_X$ and, in the limit $Y_i(0)\to\infty$, also approach the
fixed points in the space spanned by the parameters of a soft
breaking of supersymmetry. Since the set of differential equations
for $A_i(t)$ and $m_i^2(t)$ is linear, the $A$, $M_{1/2}$, and
$m_0^2$ dependence of the parameters of a soft breaking of
supersymmetry at the electroweak scale can be explicitly obtained
for universal boundary conditions. It turns out that, near the
quasi--fixed points, all $A_i(t)$ and all $\mathfrak{M}_i^2(t)$
are proportional to $M_{1/2}$ and $M_{1/2}^2$, respectively. Thus,
we have shown that, in the parameter space region considered here,
the solutions to the renormalization group equations for the
trilinear coupling constants and for some combinations of the
scalar particle masses are focused in a narrow interval within the
infrared region. Since the neutral scalar field $Y$ is not
renormalized by gauge interactions, $A_\varkappa(t)$ and
$\mathfrak{M}_\varkappa^2(t)$ are concentrated near zero;
therefore they are still dependent on the initial conditions. The
parameters $A_t(t_0)$ and $\mathfrak{M}_t^2(t_0)$ show the weakest
dependence on $A$ and $m_0^2$ because these parameters are
renormalized by strong interactions. By considering that the
quantities $\mathfrak{M}_i^2(t_0)$ are virtually independent of
the boundary conditions, we have calculated, near the quasi--fixed
points, the values of the scalar particle masses at the
electroweak scale.

In the general case of nonuniversal boundary conditions, the
solutions to the renormalization group equations within the NMSSM
for $A_i(t)$ and $\mathfrak{M}_i^2(t)$ are grouped near some
straight lines and planes in the space spanned by the parameters
of a soft breaking of supersymmetry. Moving along these lines and
surfaces as $Y_i(0)$ is increased, the trilinear coupling
constants and the above combinations of he scalar particle masses
approach quasi--fixed points. The formation of the straight lines
and planes can be traced by considering the examples of an exact
(at $Y_\lambda=0$) and an approximate (for $Y_\lambda\ne 0$)
solution. Although the approximate solution for the parameters of
a soft breaking of supersymmetry in a nonminimal SUSY model (this
solution has been obtained here for the first time) describes the
evolution of $A_\varkappa(t)$ and $\mathfrak{M}_\varkappa^2(t)$
very poorly, it leads to qualitatively correct results.

However, our analysis of the behaviour of solutions to the
renormalization group equations within the NMSSM in the regime of
strong Yukawa coupling has revealed that the $A_i(0)$ and
$\mathfrak{M}_i^2(0)$ dependence of the trilinear coupling
constants and the above combinations of the scalar particle masses
dies out quite slowly, in proportion to
$(\epsilon_t(t))^{\lambda_{\text{min}}}$, where
$\epsilon_t(t)=Y_t(t)/Y_t(0)$ and $\lambda_{\text{min}}$ is the
minimal eigenvalue of the matrix $c_{ij}r_j^0$. As a rule,
$\lambda_{\text{min}}$ is positive and much less than unity. For
example, $\lambda_{\text{min}}=3/7$ at $Y_\varkappa=0$ and
$\lambda_{\text{min}}\approx 0.0085$ at $Y_\varkappa\ne 0$. The
above is invalid only for the solutions $A_i(t)$ and
$\mathfrak{M}_i^2(t)$ that correspond to universal boundary
conditions for the parameters of a soft breaking of supersymmetry
and to the initial values of $R_{\lambda 0}=1,~R_{\varkappa 0}=0$
and $R_{\lambda 0}=3/4,~R_{\varkappa 0}=3/8$ for the Yukawa
coupling constants at the Grand Unification scale. They correspond
to quasi--fixed points of the renormalization group equations for
$Y_i(t)$. As the Yukawa coupling constants are increased, such
solutions are attracted to infrared quasi--fixed points in
proportion to $\epsilon_t(t)$.

It is precisely because of the inequality $\lambda_{\text{min}}\ll
1$ that straight lines in the $(A_t,A_\lambda,A_\varkappa)$ and
$(\mathfrak{M}_t^2,\mathfrak{M}_\lambda^2,\mathfrak{M}_\varkappa^2)$
spaces, rather than infrared quasi--fixed points, play a key role
in the analysis of the behaviour of solutions for $A_i(t)$ and
$\mathfrak{M}_i^2(t)$ in the case where
$Y_t(0),Y_\lambda(0),Y_\varkappa(0)\sim 1$. In the space spanned
by the parameters of a soft breaking of supersymmetry, these
straight lines lie in the planes near which $A_i(t)$ and
$\mathfrak{M}_i^2(t)$ are grouped in the regime of strong Yukawa
coupling at the electroweak scale. A method has been proposed in
the present study for deriving equations that describe the
straight lines and planes being discussed. The straight lines and
planes that were obtained by using this method or by fitting the
results of our numerical calculations are nearly orthogonal to the
$A_t$ and $\mathfrak{M}_t^2$ axes. This is because the constants
$A_t(t_0)$ and $\mathfrak{M}_t^2(t_0)$ are virtually independent
of the initial conditions at the scale $M_X$. On the other hand,
the parameters $A_\varkappa(t_0)$ and
$\mathfrak{M}_\varkappa^2(t_0)$ are determined, to a considerable
extent, by the boundary conditions at the scale $M_X$. At
$R_{\lambda 0}=3/4$ and $R_{\varkappa 0}=3/8$, the planes in the
$(A_t,A_\lambda,A_\varkappa)$ and
$(\mathfrak{M}_t^2,\mathfrak{M}_\lambda^2,\mathfrak{M}_\varkappa^2)$
spaces are therefore parallel to the $A_\varkappa$ and
$\mathfrak{M}_\varkappa^2$ axes.

Within the NMSSM involving a minimal set of fundamental
parameters, we have been unfortunately unable to construct, at
$Y_i(0)\gg\tilde{\alpha}_i(0)$, a self--consistent solution that,
on one hand. would lead to a heavy spectrum of SUSY particles and
which, on the other hand, would make it possible to obtain, for
the lightest Higgs boson, a mass value greater than that in the
MSSM. In order to obtain such a solution, it is necessary to
modify the nonminimal SUSY model. In the near future, we plan to
investigate the spectrum of the superpartners of observable
particles and Higgs bosons within the simplest modification of the
NMSSM such that it leads to a self--consistent solution in the
regime of strong Yukawa coupling.

\section*{Acknowledgements}

The authors are grateful to M. I. Vysotsky, D. I. Kazakov, and K.
A. Ter--Martirosyan for stimulating discussions and comments. R.
B. Nevzorov is indebted to DESY Theory Group for hospitality
extended to him.

This work was supported by the Russian Foundation for Basic
Research (RFBR), projects \#\# 98-02-17372, 98-02-17453,
00-15-96786, 00-15-96562.

\newpage

\section*{Appendix}

{\bfseries Set of the renormalization group equations of the NMSSM
and an approximate solution of it.}

The full set of one--loop renormalization group equations of the
NMSSM can be represented in the form \cite{28}

\begin{equation}
\label{A.1}
\begin{split}
\frac{d\tilde{\alpha}_i}{dt}=&-b_i\tilde{\alpha}_i^2,\\
\frac{dY_t}{dt}=&-Y_t(Y_{\lambda}+6Y_t
-\frac{16}{3}\tilde{\alpha}_3-3\tilde{\alpha}_2-\frac{13}{15}\tilde{\alpha}_1),\\
\frac{dY_{\lambda}}{dt}=&-Y_{\lambda}(4Y_{\lambda}+2Y_{\varkappa}+3Y_t
 -3\tilde{\alpha}_2-\frac{3}{5}\tilde{\alpha}_1),\\
\frac{dY_\varkappa}{dt}=&-6Y_{\varkappa}(Y_{\lambda}+Y_{\varkappa}),\\
\frac{dM_i}{dt}=&-b_i\tilde{\alpha}_iM_i,\\
\frac{dA_t}{dt}=&-Y_{\lambda} A_{\lambda}-6Y_tA_t
+\frac{16}{3}\tilde{\alpha}_3 M_3+3\tilde{\alpha}_2 M_2
+\frac{13}{15}\tilde{\alpha}_1 M_1,\\
\frac{dA_{\lambda}}{dt}=&-4Y_{\lambda} A_{\lambda} -2Y_{\varkappa}
A_{\varkappa}-3Y_t A_t+3\tilde{\alpha}_2 M_2+
\frac{3}{5}\tilde{\alpha}_1 M_1,\\
\frac{dA_\varkappa}{dt}=&-6Y_{\lambda} A_{\lambda}- 6Y_{\varkappa}
A_{\varkappa},\\
\frac{dm_1^2}{dt}=&-Y_{\lambda}(m_1^2+m_2^2+m_y^2+A_{\lambda}^2)
+3\tilde{\alpha}_2M_2^2+\frac{3}{5}\tilde{\alpha}_1M_1^2,\\
\frac{dm_2^2}{dt}=&-Y_{\lambda}(m_1^2+m_2^2+m_y^2+A_{\lambda}^2)
-3Y_t(m_2^2+m_U^2+m_Q^2+A_t^2)+{}\\ &{}+3\tilde{\alpha}_2
M_2^2+\frac{3}{5}\tilde{\alpha}_1 M_1^2,\\
\frac{dm_y^2}{dt}=&-2Y_{\lambda}(m_1^2+m_2^2+m_y^2+A_{\lambda}^2)
-2Y_{\varkappa}(3m_y^2+A_{\varkappa}^2),\\
\frac{dm_U^2}{dt}=&-2Y_t(m_2^2+m_U^2+m_Q^2+A_t^2)
+\frac{16}{3}\tilde{\alpha}_3 M_3^2+\frac{16}{15}
\tilde{\alpha}_1M_1^2,\\
\frac{dm_Q^2}{dt}=&-Y_t(m_2^2+m_U^2+m_Q^2+A_t^2)
+\frac{16}{3}\tilde{\alpha}_3 M_3^2+3\tilde{\alpha_2}
M_2^2+\frac{1}{15}\tilde{\alpha}_1 M_1^2,\\
\end{split}
\end{equation}
where the index $i$ runs through the values from $1$ to $3$,
$b_1=33/5$, $b_2=1$, $b_3=-3$,
$\tilde{\alpha}_i(t)=\alpha_i(t)/(4\pi)=g_i^2(t)/(4\pi)^2$,
$Y_t(t)=h_t^2(t)/(4\pi)^2$, $Y_\lambda(t)=\lambda^2(t)/(4\pi)^2$,
and $Y_\varkappa(t)=\varkappa^2(t)/(4\pi)^2$. The variable $t$ is
defined in a conventional way: $t=\ln(M_X^2/Q^2)$, where
$M_X=3\cdot 10^{16}\text{~GeV}$. On the right--hand sides of
(\ref{A.1}), we have discarded terms proportional to the Yukawa
coupling constants $Y_b$ and $Y_\tau$, because their contribution
is negligibly small at $\tan\beta\sim 1$. For the Yukawa and gauge
coupling constants, the set of two--loop renormalization group
equations within the NMSSM is presented in \cite{29}. Using the
results obtained from an analysis of renormalization of various
types of constants in SUSY theories \cite{31} and also the
two--loop equations for the Yukawa and the gauge coupling
constants, one can easily obtain the relevant renormalization
group equations describing the evolution of parameters of a soft
breaking of supersymmetry. Moreover, the full set of two--loop
renormalization group equations of the NMSSM can be obtained
directly from the results presented in \cite{32}.

Although an exact analytic solution to the set of one--loop
renormalization group equations (\ref{A.1}) of the NMSSM is not
known at present, an approximate solution for the parameters of a
soft breaking of supersymmetry can be constructed on the  basis of
the approximate solution for the Yukawa coupling constants
\cite{19}:
\begin{equation}
\label{A.2}
\begin{split}
Y_t(t)=&\frac{Y_t(0)E_t(t)}{(1+6Y_t(0)F_t(t))^{1/2}(1+6Y_t(0)F_t(t)+
2Y_{\lambda}(0)F_{\lambda}(t))^{1/2}},\\
Y_{\lambda}(t)=&\frac{Y_{\lambda}(0)E_{\lambda}(t)}%
{(1+6Y_t(0)F_t(t)+2Y_{\lambda}(0)F_{\lambda}(t))^{1/2}
(1+6Y_{\lambda}(0)F_{\lambda}(t))^{1/6}}\\
&{}\times\frac{1}{(1+6Y_{\lambda}(0)F_{\lambda}(t)+6Y_{\varkappa}(0)t)^{1/3}},\\
Y_{\varkappa}(t)=&\frac{Y_{\varkappa}(0)}{(1+6Y_{\lambda}(0)F_{\lambda}(t)
+6Y_{\varkappa}(0)t)},\\
E_t(t)=&\left[\frac{\tilde{\alpha}_3(t)}{\tilde{\alpha}(0)}\right]^{16/9}
\left[\frac{\tilde{\alpha}_2(t)}{\tilde{\alpha}(0)}\right]^{-3}
\left[\frac{\tilde{\alpha}_1(t)}{\tilde{\alpha}(0)}\right]^{-13/99},
\quad F_t(t)=\int\limits_0^t E_t(t')dt',\\
E_{\lambda}(t)=&\left[\frac{\tilde{\alpha}_2(t)}{\tilde{\alpha}(0)}\right]^{-3}
\left[\frac{\tilde{\alpha}_1(t)}{\tilde{\alpha}(0)}\right]^{-1/11},
\quad F_{\lambda}(t)=\int\limits_0^t E_{\lambda}(t')dt',\\
\tilde{\alpha}_i(t)=&\frac{\tilde{\alpha}(0)}
{1+b_i\tilde{\alpha}(0)t}.
\end{split}
\end{equation}
To do this, it is necessary to replace the quantities
$\tilde{\alpha}_i$ and $Y_i$ in the approximate solution
(\ref{A.2}) by $\tilde{\alpha}'_i$ and $Y'_i$:
\begin{equation}
\label{A.3}
\begin{split}
\tilde{\alpha}'_i&=\tilde{\alpha}_i\left(1+M_i\eta+\bar{M}_i
\bar{\eta}+2M_i\bar{M}_i\eta\overline{\eta}\right),\\
Y'_i&=Y_i\left(1+A_i\eta+\bar{A}_i
\bar{\eta}+2A_i\bar{A}_i\eta\bar{\eta}+\mathfrak{M}_i^2
\eta\bar{\eta}\right),\\ \eta&=\theta^2,\quad
\bar{\eta}=\bar{\theta}^2,
\end{split}
\end{equation}
where $\theta$ are Grassmann variables, and to construct an
expansion in $\eta$. This method for deducing solutions made it
possible to simplify considerably the expressions for $A_t(t)$ and
$\mathfrak{M}_t^2(t)$ within the MSSM at $\tan\beta\sim 1$
\cite{33} and to obtain and analyse approximate solutions for the
parameters of a soft SUSY breaking within the MSSM for
$\tan\beta\gg 1$ \cite{34}. In  the most general case of
nonuniversal boundary conditions, we have
\begin{equation}
\label{A.4}
\begin{split}
A_i(t)&=\sum_j e_{ij}(t)A_j(0)+f_i(t)M_{1/2},\\
\mathfrak{M}_i^2(t)&=\sum_j
\left(\tilde{a}_{ij}(t)\mathfrak{M}_j^2(0)+
\tilde{c}_{ij}(t)A_j(0)M_{1/2}\right)+\sum_{jk}\tilde{d}_{ijk}(t)A_j(0)A_k(0)+
\tilde{b}_i(t)M_{1/2}^2,
\end{split}
\end{equation}
where $e_{ij}(t)=\tilde{a}_{ij}(t)$, since the homogeneous
equations for $A_i(t)$ and $\mathfrak{M}_i^2(t)$ coincide, and
where the functions $e_{ij}(t)$, $\tilde{a}_{ij}(t)$, $f_i(t)$,
$\tilde{b}_i(t)$, $\tilde{c}_{ij}(t)$, and $\tilde{d}_{ijk}(t)$
are expressed in terms of the partial derivatives of the Yukawa
coupling constants as
\begin{equation}
\label{A.5}
\begin{split}
e_{ij}(t)&=Y_j(0)\frac{\partial\ln Y_i(t)}{\partial Y_j(0)},\qquad
f_i(t)=\tilde{\alpha}(0)\frac{\partial\ln Y_i(t)}%
{\partial\tilde{\alpha}(0)},\\
\tilde{b}_i(t)&=\frac{\partial}{\partial\tilde{\alpha}(0)}
\left[\tilde{\alpha}^2(0)\frac{\partial\ln Y_i(t)}
{\partial\tilde{\alpha}(0)}\right]=f_i(t)+
\tilde{\alpha}(0)\frac{\partial
f_i(t)}{\partial\tilde{\alpha}(0)},\\
\tilde{c}_{ij}(t)&=2\tilde{\alpha}(0)
\frac{\partial}{\partial\tilde{\alpha}(0)}\left[Y_j(0)
\frac{\partial\ln Y_i(t)}{\partial Y_j(0)}\right]=
2\tilde{\alpha}(0)\frac{\partial e_{ij}(t)}
{\partial\tilde{\alpha}(0)},\\
\tilde{d}_{ijk}(t)&=\tilde{d}_{ikj}(t)=Y_k(0)
\frac{\partial}{\partial Y_k(0)}\left[Y_j(0) \frac{\partial\ln
Y_i(t)}{\partial Y_j(0)}\right]=Y_k(0)\frac{\partial
e_{ij}(t)}{\partial Y_k(0)}.
\end{split}
\end{equation}

Substituting the above approximate solutions for the Yukawa
coupling constants within the NMSSM into (\ref{A.5}), we obtain:

\begin{eqnarray}
e_{\lambda\lambda}(t)&=&1-\frac{2Y_\lambda(0)F_{\lambda}(t)}
{1+6Y_\lambda(0)F_\lambda(t)+6Y_\varkappa(0)t}
-\frac{Y_\lambda(0)F_\lambda(t)}{1+6Y_\lambda(0)F_\lambda(t)}
\nonumber \\ &&{}-\frac{Y_\lambda(0)F_\lambda(t)}
 {1+6Y_t(0)F_t(t)+2Y_\lambda(0)F_\lambda(t)}, \nonumber \\
e_{\lambda\varkappa}(t)&=&-\frac{2Y_\varkappa(0)t}
 {1+6Y_\lambda(0)F_\lambda(t)+6Y_\varkappa(0)t}, \nonumber \\
e_{\lambda t}(t)&=&-\frac{3Y_t(0)F_t(t)}
 {1+6Y_t(0)F_t(t)+2Y_\lambda(0)F_\lambda(t)}, \nonumber \\
e_{\varkappa\lambda}(t)&=&-\frac{6Y_\lambda(0)F_\lambda(t)}
 {1+6Y_\lambda(0)F_\lambda(t)+6Y_\varkappa(0)t}, \nonumber \\
e_{\varkappa\varkappa}(t)&=&1-\frac{6Y_\varkappa(0)t}
 {1+6Y_\lambda(0)F_\lambda(t)+6Y_\varkappa(0)t}, \nonumber \\
e_{\varkappa t}(t)&=&e_{t\varkappa}(t)=0, \label{A.6} \\
e_{t\lambda}(t)&=&-\frac{Y_\lambda(0)F_\lambda(t)}
 {1+6Y_t(0)F_t(t)+2Y_\lambda(0)F_\lambda(t)}, \nonumber \\
e_{tt}(t)&=&1-\frac{3Y_t(0)F_t(t)}{1+6Y_t(0)F_t(t)}
 -\frac{3Y_t(0)F_t(t)}{1+6Y_t(0)F_t(t)+2Y_\lambda(0)F_\lambda(t)}; \nonumber \\
f_\lambda(t)&=&\frac{tE'_\lambda(t)}{E_\lambda(t)}
 -\frac{Y_\lambda(0)(tE_\lambda(t)-F_\lambda(t))}
 {1+6Y_\lambda(0)F_\lambda(t)}
 -\frac{2Y_\lambda(0)(tE_\lambda(t)-F_\lambda(t))}
 {1+6Y_\lambda(0)F_\lambda(t)+6Y_\varkappa(0)t} \nonumber \\
 &&{}-\frac{3Y_t(0)(tE_t(t)-F_t(t))+Y_\lambda(0)(tE_\lambda(t)-
 F_\lambda(t))}{1+6Y_t(0)F_t(t)+2Y_\lambda(0)F_\lambda(t)}, \nonumber \\
f_\varkappa(t)&=&-\frac{6Y_\lambda(0)(tE_\lambda(t)-F_\lambda(t))}
 {1+6Y_\lambda(0)F_\lambda(t)+6Y_\varkappa(0)t}, \nonumber \\
f_t(t)&=&\frac{tE'_t(t)}{E_t(t)}
 -\frac{3Y_t(0)(tE_t(t)-F_t(t))}{1+6Y_t(0)F_t(t)} \nonumber \\
 &&{}-\frac{3Y_t(0)(tE_t(t)-F_t(t))+Y_\lambda(0)(tE_\lambda(t)-
 F_\lambda(t))}{1+6Y_t(0)F_t(t)+2Y_\lambda(0)F_\lambda(t)};
 \nonumber
\end{eqnarray}

\begin{eqnarray*}
\tilde b_\lambda(t)&=&\frac{d}{dt}\left(t^2\frac{E'_\lambda(t)}
 {E_\lambda(t)}\right)-\frac{Y_\lambda(0)t^2E'_\lambda(t)}
 {1+6Y_\lambda(0)F_\lambda(t)}
 +\frac{6Y^2_\lambda(0)(tE_\lambda(t)-F_\lambda(t))^2}
 {(1+6Y_\lambda(0)F_\lambda(t))^2} \\
 &&{}-\frac{3Y_t(0)t^2E'_t(t)+Y_\lambda(0)t^2E'_\lambda(t)}
 {1+6Y_t(0)F_t(t)+2Y_\lambda(0)F_\lambda(t)}
 -\frac{2Y_\lambda(0)t^2E'_\lambda(t)}
 {1+6Y_\lambda(0)F_{\lambda}(t)+6Y_\varkappa(0)t} \\
 &&{}+\frac{2(3Y_t(0)(tE_t(t)-F_t(t))+Y_\lambda(0)(tE_\lambda(t)-F_\lambda(t)))^2}
 {(1+6Y_t(0)F_t(t)+2Y_\lambda(0)F_\lambda(t))^2} \\
 &&{}+\frac{12Y_\lambda^2(0)(tE_\lambda(t)-F_\lambda(t))^2}
 {(1+6Y_\lambda(0)F_\lambda(t)+6Y_\varkappa(0)t)^2},\\
\tilde b_\varkappa(t)&=&-\frac{6Y_\lambda(0)t^2E'_\lambda(t)}
 {1+6Y_\lambda(0)F_\lambda(t)+6Y_\varkappa(0)t}
 +\frac{36Y^2_\lambda(0)(tE_\lambda(t)-F_\lambda(t))^2}
 {(1+6Y_\lambda(0)F_\lambda(t)+6Y_\varkappa(0)t)^2},\\
\tilde b_t(t)&=&\frac{d}{dt}\left(t^2\frac{E'_t(t)}
{E_t(t)}\right)
 -\frac{3Y_t(0)t^2E'_t(t)}{1+6Y_t(0)F_t(t)}
 +\frac{18Y_t^2(0)(tE_t(t)-F_t(t))^2}{(1+6Y_t(0)F_t(t))^2} \\
 &&{}+\frac{2(3Y_t(0)(tE_t(t)-F_t(t))+Y_\lambda(0)(tE_\lambda(t)-F_\lambda(t)))^2}
 {(1+6Y_t(0)F_t(t)+2Y_\lambda(0)F_\lambda(t))^2} \\
 &&{}-\frac{3Y_t(0)t^2E'_t(t)+Y_\lambda(0)t^2E'_\lambda(t)}
 {1+6Y_t(0)F_t(t)+2Y_\lambda(0)F_\lambda(t)}; \\
\tilde c_{\lambda\lambda}(t)&=&-\frac
 {4Y_\lambda(0)(tE_\lambda(t)-F_\lambda(t))(1+6Y_\varkappa(0)t)}
 {(1+6Y_\lambda(0)F_\lambda(t)+6Y_\varkappa(0)t)^2}
 -\frac{2Y_\lambda(0)(tE_\lambda(t)-F_\lambda(t))}
 {(1+6Y_\lambda(0)F_\lambda(t))^2}\\
 &&{}+12\frac{Y_t(0)Y_\lambda(0)F_\lambda(t)(tE_t(t)-F_t(t))}
 {(1+6Y_t(0)F_t(t)+2Y_\lambda(0)F_\lambda(t))^2}\\
 &&{}-2\frac{Y_\lambda(0)(tE_\lambda(t)-F_\lambda(t))(1+6Y_t(0)F_t(t))}
 {(1+6Y_t(0)F_t(t)+2Y_\lambda(0)F_\lambda(t))^2}, \hspace*{45mm} \text{(\ref{A.6})} \\
\tilde c_{\lambda\varkappa}(t)&=&\frac
 {24Y_\varkappa(0)Y_\lambda(0)(tE_\lambda(t)-F_\lambda(t))t}
 {(1+6Y_\lambda(0)F_\lambda(t)+6Y_\varkappa(0)t)^2}, \\
\tilde c_{\lambda t}(t)&=&12\frac%
 {Y_t(0)Y_\lambda(0)F_t(t)(tE_\lambda(t)-F_\lambda(t))}
 {(1+6Y_t(0)F_t(t)+2Y_\lambda(0)F_\lambda(t))^2}\\
 &&{}-6\frac{Y_t(0)(tE_t(t)-F_t(t))(1+2Y_\lambda(0)F_\lambda(t))}
 {(1+6Y_t(0)F_t(t)+2Y_\lambda(0)F_\lambda(t))^2},\\
\tilde c_{\varkappa\lambda}(t)&=&-12\frac%
 {Y_\lambda(0)(tE_\lambda(t)-F_\lambda(t))(1+6Y_\varkappa(0)t)}
 {(1+6Y_\lambda(0)F_\lambda(t)+6Y_\varkappa(0)t)^2},\\
\tilde c_{\varkappa\varkappa}(t)&=&\frac%
 {72Y_\varkappa(0)Y_\lambda(0)(tE_\lambda(t)-F_\lambda(t))t}
 {(1+6Y_\lambda(0)F_\lambda(t)+6Y_\varkappa(0)t)^2},\\
\tilde c_{\varkappa t}(t)&=&\tilde c_{t\varkappa}(t)=0,\\
\tilde c_{t\lambda}(t)&=&12\frac%
 {Y_t(0)Y_\lambda(0)F_\lambda(tE_t(t)-F_t(t))}
 {(1+6Y_t(0)F_t(t)+2Y_\lambda(0)F_\lambda(t))^2}\\
 &&{}-2\frac{Y_\lambda(0)(tE_\lambda(t)-F_\lambda(t))(1+6Y_t(0)F_t(t))}
 {(1+6Y_t(0)F_t(t)+2Y_\lambda(0)F_\lambda(t))^2},\\
\tilde c_{tt}(t)&=&-\frac{6Y_t(0)(tE_t(t)-F_t(t))}
 {(1+6Y_t(0)F_t(t))^2}+
 12\frac{Y_t(0)Y_\lambda(0)F_t(t)(tE_\lambda(t)-F_\lambda(t))}
 {(1+6Y_t(0)F_t(t)+2Y_\lambda(0)F_\lambda(t))^2}\\
&&{}-6\frac{Y_t(0)(tE_t(t)-F_t(t))(1+2Y_\lambda(0)F_\lambda(t))}
 {(1+6Y_t(0)F_t(t)+2Y_\lambda(0)F_\lambda(t))^2};
\end{eqnarray*}

\begin{eqnarray*}
\tilde{d}_{\lambda\lambda\lambda}(t)&=&-\frac
 {2Y_\lambda(0)F_\lambda(t)(1+6Y_\varkappa(0)t)}
 {(1+6Y_\lambda(0)F_\lambda(t)+6Y_\varkappa(0)t)^2}
 -\frac{Y_\lambda(0)F_\lambda(t)}
 {(1+6Y_\lambda(0)F_\lambda(t))^2}\\
 &&{}-\frac{Y_\lambda(0)F_\lambda(t)(1+6Y_t(0)F_t(t))}
 {(1+6Y_t(0)F_t(t)+2Y_\lambda(0)F_\lambda(t))^2},\\
\tilde{d}_{\lambda\lambda\varkappa}(t)&=&\tilde{d}_{\lambda\varkappa\lambda}(t)=\frac
 {12Y_\varkappa(0)Y_\lambda(0)tF_\lambda(t)}
 {(1+6Y_\lambda(0)F_\lambda(t)+6Y_\varkappa(0)t)^2},\\
\tilde{d}_{\lambda\lambda t}(t)&=&\tilde{d}_{\lambda
t\lambda}(t)=\frac
 {6Y_t(0)Y_\lambda(0)F_t(t)F_\lambda(t)}
 {(1+6Y_t(0)F_t(t)+2Y_\lambda(0)F_\lambda(t))^2},\\
\tilde{d}_{\lambda\varkappa\varkappa}(t)&=&-\frac
 {2Y_\varkappa(0)t(1+6Y_\lambda(0)F_\lambda(t))}
 {(1+6Y_\lambda(0)F_\lambda(t)+6Y_\varkappa(0)t)^2},\\
\tilde{d}_{\lambda tt}(t)&=&-\frac
 {3Y_t(0)F_t(t)(1+2Y_\lambda(0)F_\lambda(t))}
 {(1+6Y_t(0)F_t(t)+2Y_\lambda(0)F_\lambda(t))^2},\\
\tilde{d}_{\varkappa\lambda\lambda}(t)&=&-\frac
 {6Y_\lambda(0)F_\lambda(t)(1+6Y_\varkappa(0)t)}
 {(1+6Y_\lambda(0)F_\lambda(t)+6Y_\varkappa(0)t)^2}, \hspace*{60mm} \text{(\ref{A.6})} \\
\tilde{d}_{\varkappa\lambda\varkappa}(t)&=&\tilde{d}_{\varkappa\varkappa\lambda}(t)=\frac
 {36Y_\lambda(0)Y_\varkappa(0)F_\lambda(t) t}
 {(1+6Y_\lambda(0)F_\lambda(t)+6Y_\varkappa(0)t)^2},\\
\tilde{d}_{\varkappa\varkappa\varkappa}(t)&=&-\frac
 {6Y_\varkappa(0)t(1+6Y_\lambda(0)F_\lambda(t))}
 {(1+6Y_\lambda(0)F_\lambda(t)+Y_\varkappa(0)t)^2},\\
\tilde{d}_{t\lambda\lambda}(t)&=&-\frac
 {Y_\lambda(0)F_\lambda(t)(1+6Y_t(0)F_t(t))}
 {(1+6Y_t(0)F_t(t)+2Y_\lambda(0)F_\lambda(t))^2},\\
\tilde{d}_{t\lambda t}(t)&=&\tilde{d}_{tt\lambda}(t)=\frac
 {6Y_t(0)Y_\lambda(0)F_t(t)F_\lambda(t)}
 {(1+6Y_t(0)F_t(t)+2Y_\lambda(0)F_\lambda(t))^2},\\
\tilde{d}_{ttt}(t)&=&-\frac{3Y_t(0)F_t(t)}
 {(1+6Y_t(0)F_t(t))^2}-\frac
 {3Y_t(0)F_t(t)(1+2Y_\lambda(0)F_\lambda(t))}
 {(1+6Y_t(0)F_t(t)+2Y_\lambda(0)F_\lambda(t))^2},\\
\tilde{d}_{\lambda\varkappa t}(t)&=&\tilde{d}_{\lambda
t\varkappa}(t)=\tilde{d}_{\varkappa\lambda
t}(t)=\tilde{d}_{\varkappa\varkappa t}(t)=\tilde{d}_{\varkappa
t\lambda}(t)=\tilde{d}_{\varkappa t\varkappa}(t)=0,\\
\tilde{d}_{\varkappa
tt}(t)&=&\tilde{d}_{t\lambda\varkappa}(t)=\tilde{d}_{t\varkappa\lambda}(t)=
\tilde{d}_{t\varkappa\varkappa}(t)=\tilde{d}_{t\varkappa
t}(t)=\tilde{d}_{tt\varkappa}(t)=0.
\end{eqnarray*}

At $Y_\lambda(0)=0$, the solution given by (\ref{A.2}) and
(\ref{A.6}) coincides with the exact analytic solution of the set
of renormalization group equations (\ref{A.1}). Summing the above
expressions for $e_{ij}(t)$, $\tilde{c}_{ij}(t)$, and
$\tilde{d}_{ijk}(t)$,
\[
e_i(t)=\tilde{a}_i(t)=\sum_j e_{ij}(t),\quad \tilde{c}_i(t)=
\sum_j\tilde{c}_{ij}(t),\quad
\tilde{d}_i(t)=\sum_{jk}\tilde{d}_{ijk}(t),
\]
we find the functions $e_i(t)$, $\tilde{c}_i(t)$, and
$\tilde{d}_i(t)$, which determine the dependence of the parameters
of a soft SUSY breaking on $A$, $M_{1/2}$, and $m_0^2$ for the
universal boundary conditions.

\newpage

\begin{center}

{\bfseries Table.} Values of the functions $e_i(t_0)$, $f_i(t_0)$,
$a_i(t_0)$, $b_i(t_0)$, $c_i(t_0)$, and $d_i(t_0)$ versus a choice
of $\varkappa^2(0)$, $\lambda^2(0)$, and $h_t^2(0)$.

\vspace*{5mm}

\begin{tabular}{|c|c|c|c|c|c|c|c|c|}
\hline $\varkappa^2(0)$&0&0&0&0&0&4&10&10\\ \hline
$\lambda^2(0)$&2&4&10&10&10&4&10&10\\ \hline
$h_t^2(0)$&10&10&10&4&2&10&10&4\\ \hline
$e_t(t_0)$&0.0011&0.0024&0.0082&0.0310&0.0651&0.0070&0.0102&0.0308\\
\hline
$e_{\lambda}(t_0)$&0.2127&0.1192&0.0260&-0.0227&-0.0544&0.0661&0.0113&
-0.0232\\ \hline
$e_{\varkappa}(t_0)$&0.5066&0.3451&0.1804&0.0774&-0.0062&0.0430&-0.0136&
-0.0528\\ \hline
$f_{t}(t_0)$&1.7196&1.7406&1.7698&1.8352&1.9272&1.7254&1.7489&1.8119\\
\hline
$f_{\lambda}(t_0)$&-0.4414&-0.4295&-0.4161&-0.3976&-0.3722&-0.4386&-0.4262&
-0.4046\\ \hline
$f_{\varkappa}(t_0)$&0.0173&0.0195&0.0175&-0.0069&-0.0422&0.0232&0.0273&
0.0138\\ \hline
$a_1(t_0)$&0.7533&0.6725&0.5902&0.5387&0.4969&0.7632&0.6882&0.6247\\
\hline
$a_2(t_0)$&-0.6217&-0.6601&-0.6927&-0.6842&-0.6539&-0.6079&-0.6406&-0.6415\\
\hline
$a_y(t_0)$&0.5066&0.3451&0.1804&0.0774&-0.0062&0.0430&-0.0136&-0.0528\\
\hline
$a_Q(t_0)$&0.5417&0.5558&0.5724&0.5924&0.6164&0.5430&0.5571&0.5779\\
\hline
$a_U(t_0)$&0.0833&0.1116&0.1448&0.1847&0.2328&0.0859&0.1141&0.1558\\
\hline
$b_1(t_0)$&0.5557&0.5665&0.5761&0.5757&0.5703&0.5538&0.5627&0.5646\\
\hline
$b_2(t_0)$&-3.0399&-3.0070&-2.9664&-2.9044&-2.8175&-3.0352&-3.0014&-2.9402\\
\hline
$b_y(t_0)$&0.0724&0.0939&0.1131&0.1122&0.1015&0.0695&0.0907&0.1004\\
\hline
$b_Q(t_0)$&5.3129&5.3202&5.3305&5.3514&5.3821&5.3150&5.3234&5.3432\\
\hline
$b_U(t_0)$&3.6951&3.7099&3.7305&3.7722&3.8336&3.6995&3.7162&3.7557\\
\hline
$c_1(t_0)$&0.0034&0.0036&0.0040&0.0087&0.0145&0.0024&0.0034&0.0080\\
\hline
$c_2(t_0)$&-0.0085&-0.0096&-0.0165&-0.0543&-0.1080&-0.0159&-0.0197&-0.0549\\
\hline
$c_y(t_0)$&0.0068&0.0072&0.0079&0.0173&0.0290&0.0052&0.0056&0.0127\\
\hline
$c_Q(t_0)$&-0.0040&-0.0044&-0.0068&-0.0210&-0.0409&-0.0061&-0.0077&-0.0210\\
\hline
$c_U(t_0)$&-0.0079&-0.0088&-0.0137&-0.0420&-0.0817&-0.0122&-0.0154&-0.0419\\
\hline
$d_1(t_0)$&-0.0186&-0.0144&-0.0055&-0.0008&0.0023&-0.0063&-0.0029&
-0.0010\\ \hline
$d_2(t_0)$&-0.0143&-0.0114&-0.0069&-0.0122&-0.0217&-0.0076&-0.0062&-0.0129\\
\hline
$d_y(t_0)$&-0.0372&-0.0288&-0.0109&-0.0016&0.0045&-0.0196&-0.0034&0.0044\\
\hline
$d_Q(t_0)$&0.0014&0.0010&-0.0005&-0.0038&-0.0080&-0.0004&-0.0011&
-0.0039\\ \hline
$d_U(t_0)$&0.0029&0.0020&-0.0010&-0.0076&-0.0160&-0.0008&-0.0022&
-0.0079\\ \hline
$\tilde{a}_t(t_0)$&0.0033&0.0073&0.0245&0.0929&0.1953&0.0210&0.0305&0.0923\\
\hline
$\tilde{a}_{\lambda}(t_0)$&0.6382&0.3575&0.0779&-0.0680&-0.1631&0.1983&
0.0340&-0.0695\\ \hline
$\tilde{a}_{\varkappa}(t_0)$&1.5199&1.0352&0.5411&0.2323&-0.0185&0.1290&-0.0407&
-0.1583\\ \hline
$\tilde{b}_t(t_0)$&5.9680&6.0231&6.0947&6.2192&6.3981&5.9794&6.0382&6.1588\\
\hline
$\tilde{b}_{\lambda}(t_0)$&-2.4118&-2.3466&-2.2771&-2.2165&-2.1457&-2.4119&
-2.3479&-2.2752\\ \hline
$\tilde{b}_{\varkappa}(t_0)$&0.2172&0.2817&0.3394&0.3367&0.3045&0.2085&0.2722&
0.3011\\ \hline
$\tilde{c}_t(t_0)$&-0.0204&-0.0228&-0.0370&-0.1173&-0.2306&-0.0342&-0.0427&
-0.1178\\ \hline
$\tilde{c}_{\lambda}(t_0)$&0.0017&0.0012&-0.0047&-0.0283&-0.0645&-0.0083&
-0.0107&-0.0342\\ \hline
$\tilde{c}_{\varkappa}(t_0)$&0.0204&0.0216&0.0238&0.0520&0.0870&0.0156&0.0166&
0.0380\\ \hline
$\tilde{d}_t(t_0)$&-0.0099&-0.0084&-0.0084&-0.0235&0.0457&-0.0088&-0.0095&
-0.0247\\ \hline
$\tilde{d}_{\lambda}(t_0)$&-0.0700&-0.0547&-0.0233&-0.0146&-0.0150&-0.0334&
-0.0125&-0.0096\\ \hline
$\tilde{d}_{\varkappa}(t_0)$&-0.1115&-0.0865&-0.0327&-0.0049&0.0135&-0.0587&
-0.0103&0.0132\\ \hline
\end{tabular}

\end{center}

\newpage

\section*{Figure captions}

{\bfseries Fig.~1.} Evolution of the trilinear couplings $A_t(t)$
and $A_\lambda(t)$ from the Grand Unification scale ($t=0$) to the
electroweak scale ($t=t_0$) at $\varkappa^2=0$ and
$\lambda^2(0)=h_t^2(0)=10$ for universal initial conditions.
Parameter $A$ varies in the range $-M_{1/2}\le A\le M_{1/2}$.\\

{\bfseries Fig.~2.} Evolution of the combinations of masses
$\mathfrak{M}_t^2(t)$ and $\mathfrak{M}_\lambda^2(t)$ from the
Grand Unification scale ($t=0$) to the electroweak scale ($t=t_0$)
at $\varkappa^2=0$ and $\lambda^2(0)=h_t^2(0)=10$ for universal
initial conditions. Parameter $m_0^2$ varies in the range $0\le
m_0^2 \le M_{1/2}^2$.\\

{\bfseries Fig.~3.} Evolution of the trilinear couplings $A_t(t)$,
$A_\lambda(t)$, and $A_\varkappa(t)$ from the Grand Unification
scale ($t=0$) to the electroweak scale ($t=t_0$) at
$h_t^2(0)=\lambda^2(0)=\varkappa^2(0)=10$ for universal initial
conditions. Parameter $A$ varies in the range $-M_{1/2}\le A\le
M_{1/2}$.\\

{\bfseries Fig.~4.} Evolution of the combinations of masses
$\mathfrak{M}_t^2(t)$, $\mathfrak{M}_\lambda^2(t)$, and
$\mathfrak{M}_\varkappa^2(t)$ from the Grand Unification scale
($t=0$) to the electroweak scale ($t=t_0$) at
$h_t^2(0)=\lambda^2(0)=\varkappa^2(0)=10$ for universal initial
conditions. Parameter $m_0^2$ varies in the range $0\le m_0^2\le
M_{1/2}^2$.\\

{\bfseries Fig.~5.} Boundary conditions for the renormalization
group equations of the NMSSM at the Grand Unification scale
($t=0$) at $\varkappa^2=0$ and $h_t^2(0)=\lambda^2(0)=20$
uniformly distributed in the $(A_t/M_{1/2},A_\lambda/M_{1/2})$
plane -- Fig. 5a, and the corresponding values of the trilinear
couplings at the electroweak scale ($t=t_0$) -- Fig. 5b. The
straight line in Fig. 5b is a fit of the values
$(A_t(t_0),A_\lambda(t_0))$.\\

{\bfseries Fig.~6.} Boundary conditions for the renormalization
group equations of the NMSSM at the Grand Unification scale
($t=0$) at $\varkappa^2=0$, $h_t^2(0)=\lambda^2(0)=20$, and
$A_t(0)=A_\lambda(0)=0$ uniformly distributed in the
$(\mathfrak{M}_t^2/M_{1/2}^2,\mathfrak{M}_\lambda^2/M_{1/2}^2)$
plane -- Fig. 6a, and the corresponding values of the trilinear
couplings at the electroweak scale ($t=t_0$) -- Fig. 6b. The
straight line in Fig. 6b is a fit of the values
$(\mathfrak{M}_t^2(t_0),\mathfrak{M}_\lambda^2(t_0))$.\\

{\bfseries Fig.~7.} Planes in the parameter spaces
$(A_t/M_{1/2},A_\lambda/M_{1/2},A_\varkappa/M_{1/2})$ -- Fig. 7a.,
and
$(\mathfrak{M}_t^2/M_{1/2}^2,\mathfrak{M}_\lambda^2/M_{1/2}^2,\mathfrak{M}_\varkappa^2/M_{1/2}^2)$
-- Fig. 7b. The shaded parts of the planes correspond to the
regions near which the solutions at $h_t^2(0)=16$,
$\lambda^2(0)=12$, and $\varkappa^2(0)=6$ are concentrated. The
initial values $A_i(0)$ and $\mathfrak{M}_i^2(0)$ vary in the
ranges $-M_{1/2}\le A\le M_{1/2}$ and $0\le\mathfrak{M}_i^2(0)\le
3M_{1/2}^2$, respectively.\\

{\bfseries Fig.~8.} Planes in the parameter spaces
$(A_t/M_{1/2},A_\lambda/M_{1/2},A_\varkappa/M_{1/2})$ -- Fig. 8a,
and
$(\mathfrak{M}_t^2/M_{1/2}^2,\mathfrak{M}_\lambda^2/M_{1/2}^2,\mathfrak{M}_\varkappa^2/M_{1/2}^2)$
-- Fig. 8b. The shaded parts of the planes correspond to the
regions near which the solutions at $h_t^2(0)=32$,
$\lambda^2(0)=24$, and $\varkappa^2(0)=12$ are concentrated. The
initial values $A_i(0)$ and $\mathfrak{M}_i^2(0)$ vary in the
ranges $-M_{1/2}\le A\le M_{1/2}$ and $0\le\mathfrak{M}_i^2(0)\le
3M_{1/2}^2$, respectively.\\

{\bfseries Fig.~9.} Set of points in planes
$(0.0223(A_{\varkappa}/M_{1/2})+0.1278(A_{\lambda}/M_{1/2}),~A_t/M_{1/2})$
-- Fig. 9a, and
$(0.0223(\mathfrak{M}^2_{\varkappa}/M_{1/2}^2)+0.1278(\mathfrak{M}^2_{\lambda}/M_{1/2}^2),
~\mathfrak{M}^2_t/M^2_{1/2})$ -- Fig. 9b, corresponding to the
values of parameters of soft SUSY breaking for $h_t^2(0)=32$,
$\lambda^2(0)=24$, $\varkappa^2(0)=12$, and for a uniform
distribution of the boundary conditions in the parameter spaces
$(A_t,A_\lambda,A_\varkappa)$ and $(\mathfrak{M}^2_t,
\mathfrak{M}^2_{\lambda},\mathfrak{M}^2_{\varkappa})$. The initial
values $A_i(0)$ and $\mathfrak{M}_i^2(0)$ vary in the ranges
$-M_{1/2}\le A\le M_{1/2}$ and $0\le\mathfrak{M}_i^2(0)\le
3M_{1/2}^2$, respectively. The straight lines in Figs. 9a and 9b
correspond to the planes in Figs. 8a and 8b, respectively.

\newpage

\begin{center}

\includegraphics[width=160mm,keepaspectratio=true]{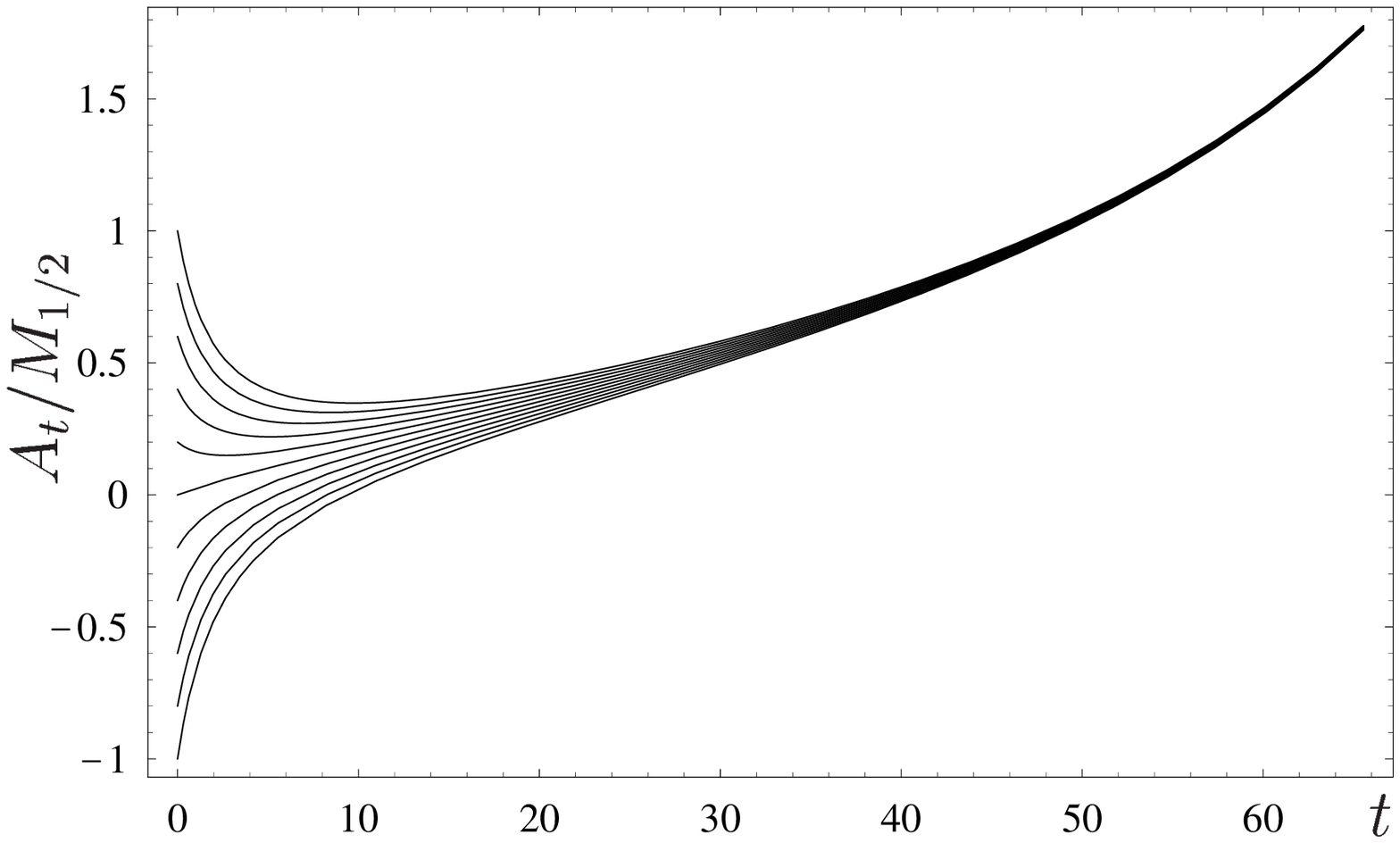}

\vspace{4mm}\hspace*{17mm}{\large\bfseries Fig.1a.}

\vspace{20mm}\includegraphics[width=160mm,keepaspectratio=true]{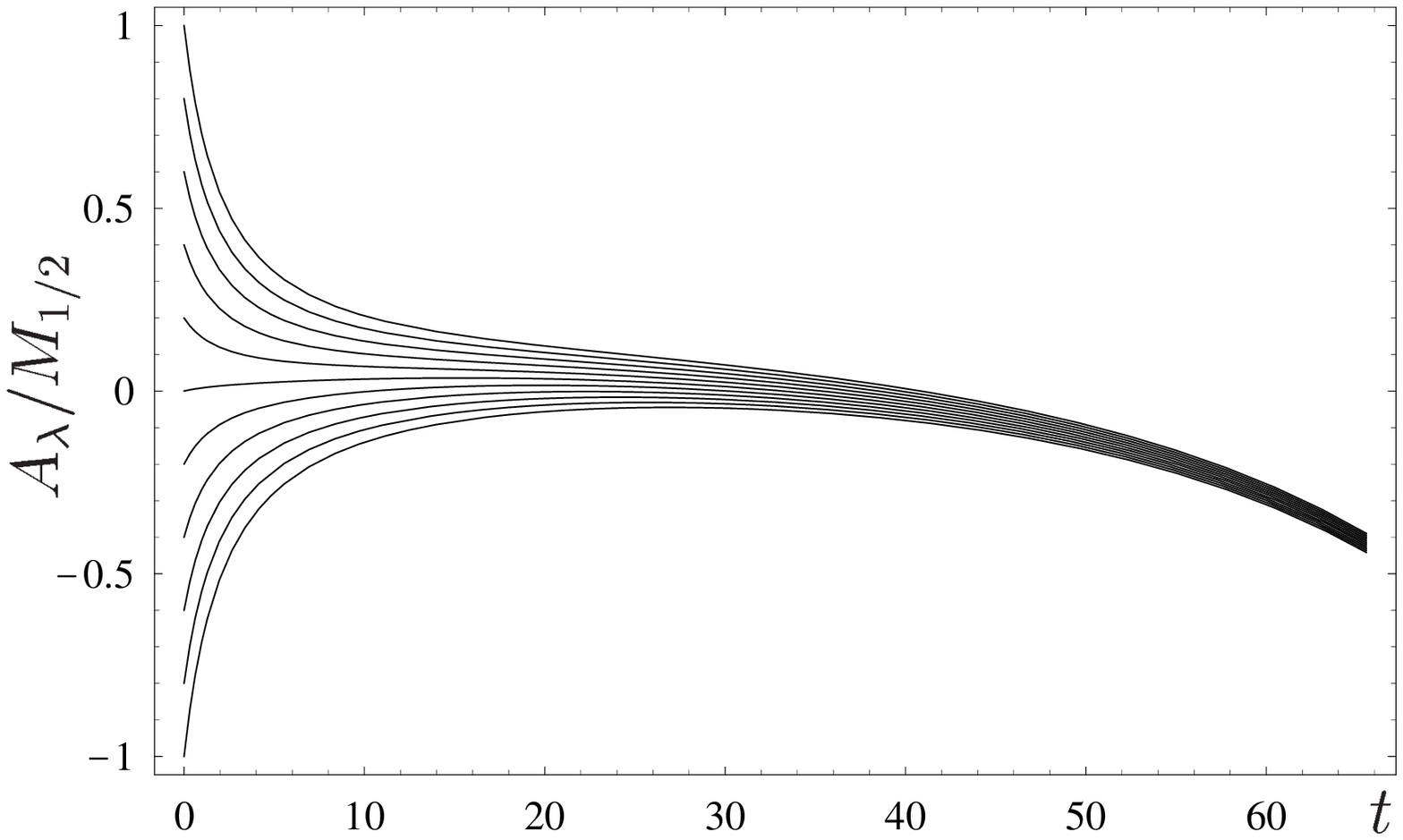}

\vspace{4mm}\hspace*{17mm}{\large\bfseries Fig.1b.}

\end{center}

\newpage

\begin{center}

\includegraphics[width=160mm,keepaspectratio=true]{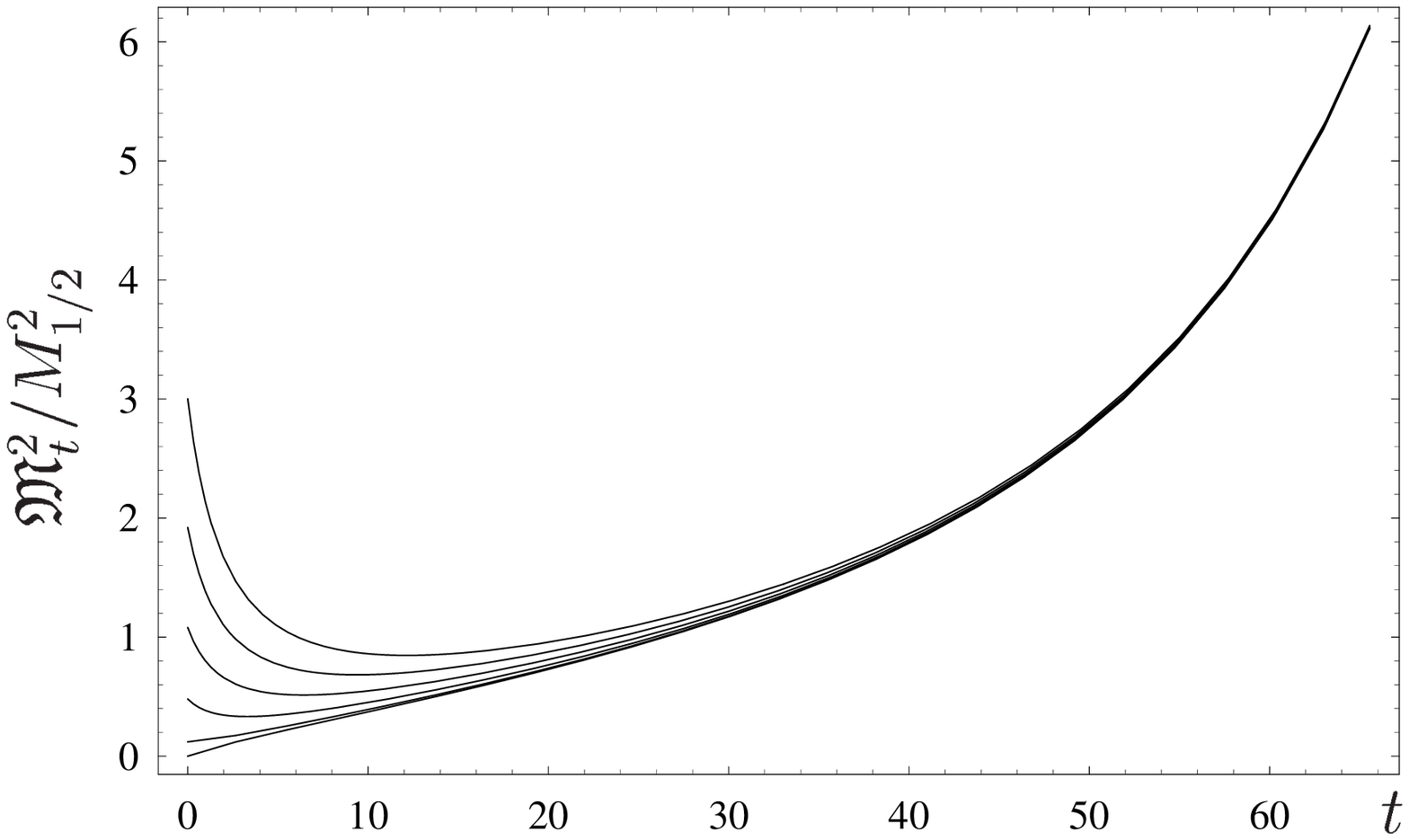}

\vspace{4mm}\hspace*{17mm}{\large\bfseries Fig.2a.}

\vspace{20mm}\includegraphics[width=160mm,keepaspectratio=true]{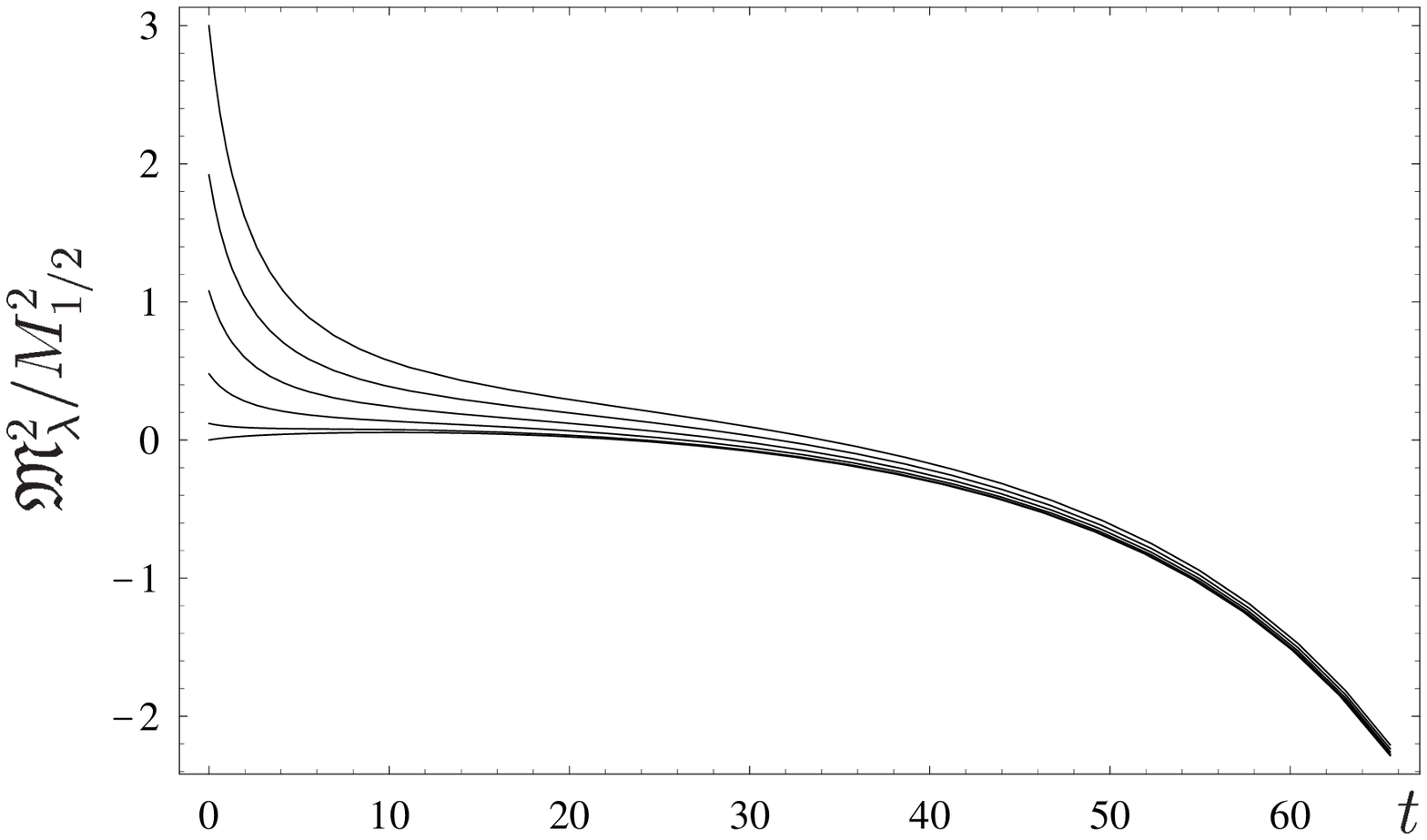}

\vspace{4mm}\hspace*{17mm}{\large\bfseries Fig.2b.}

\end{center}

\newpage

\begin{center}

\vspace*{-15mm}\includegraphics[height=69mm,keepaspectratio=true]{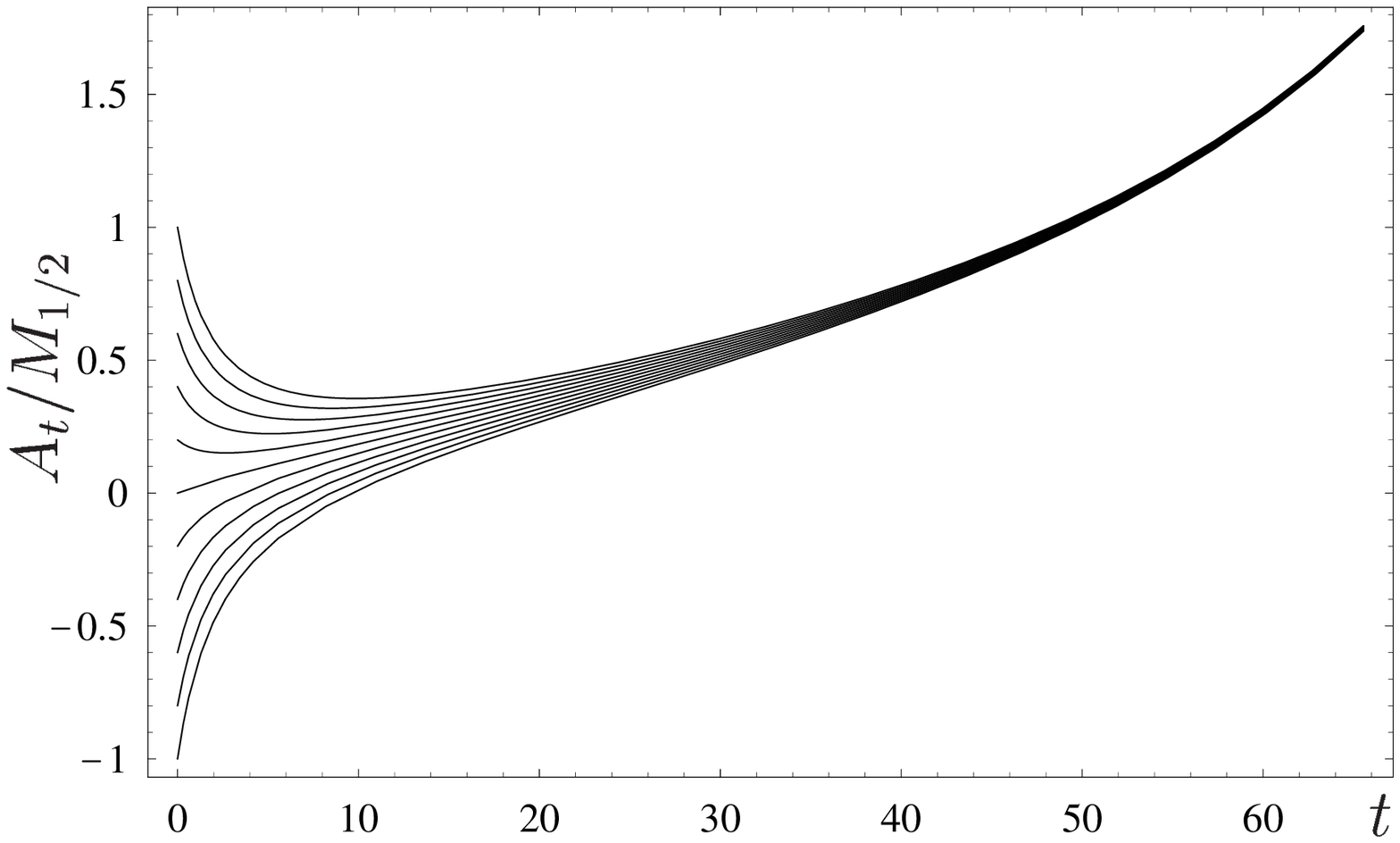}

\vspace{0mm}\hspace*{12mm}{\large\bfseries Fig.3a.}

\vspace{8mm}\includegraphics[height=69mm,keepaspectratio=true]{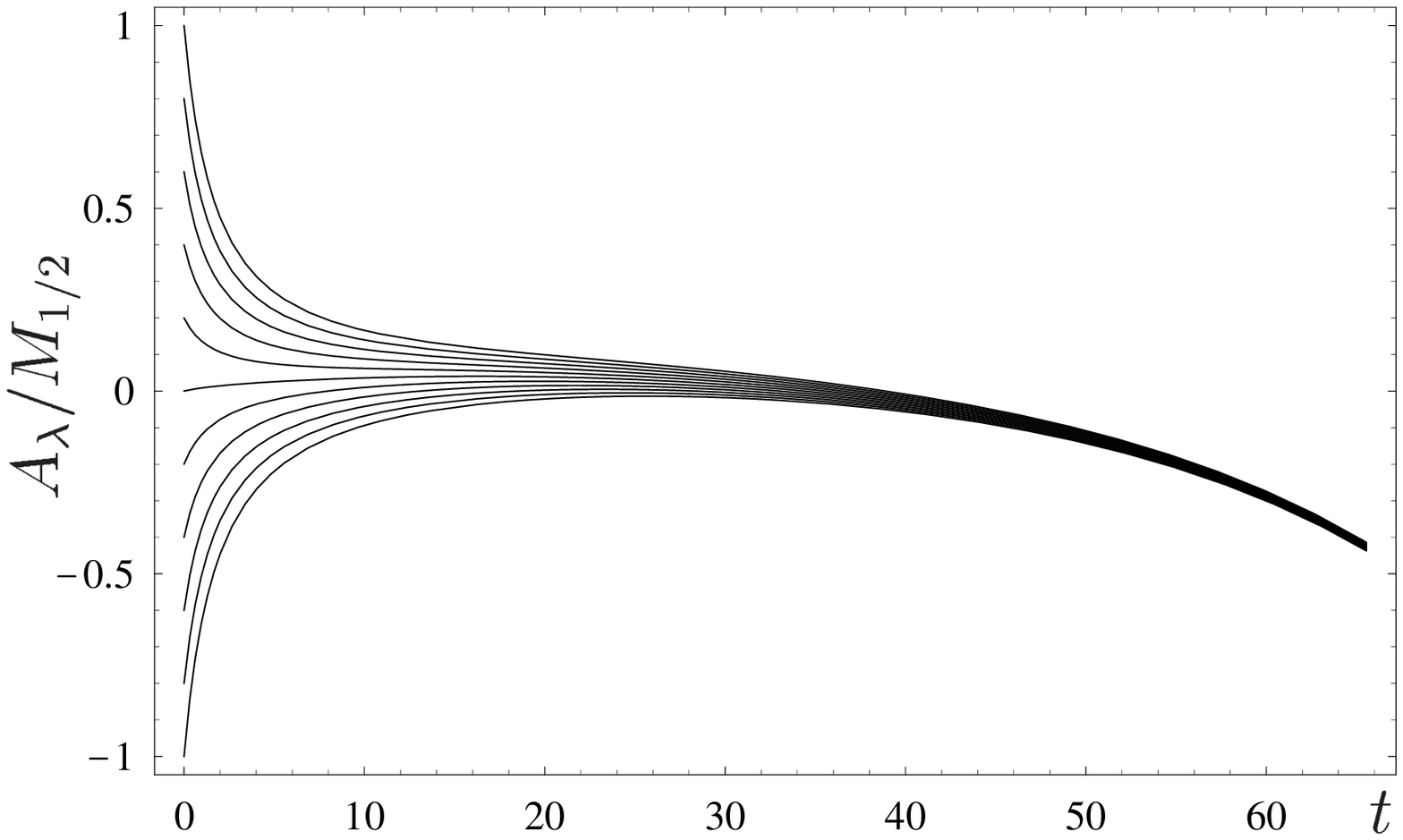}

\vspace{0mm}\hspace*{12mm}{\large\bfseries Fig.3b.}

\vspace{8mm}\includegraphics[height=69mm,keepaspectratio=true]{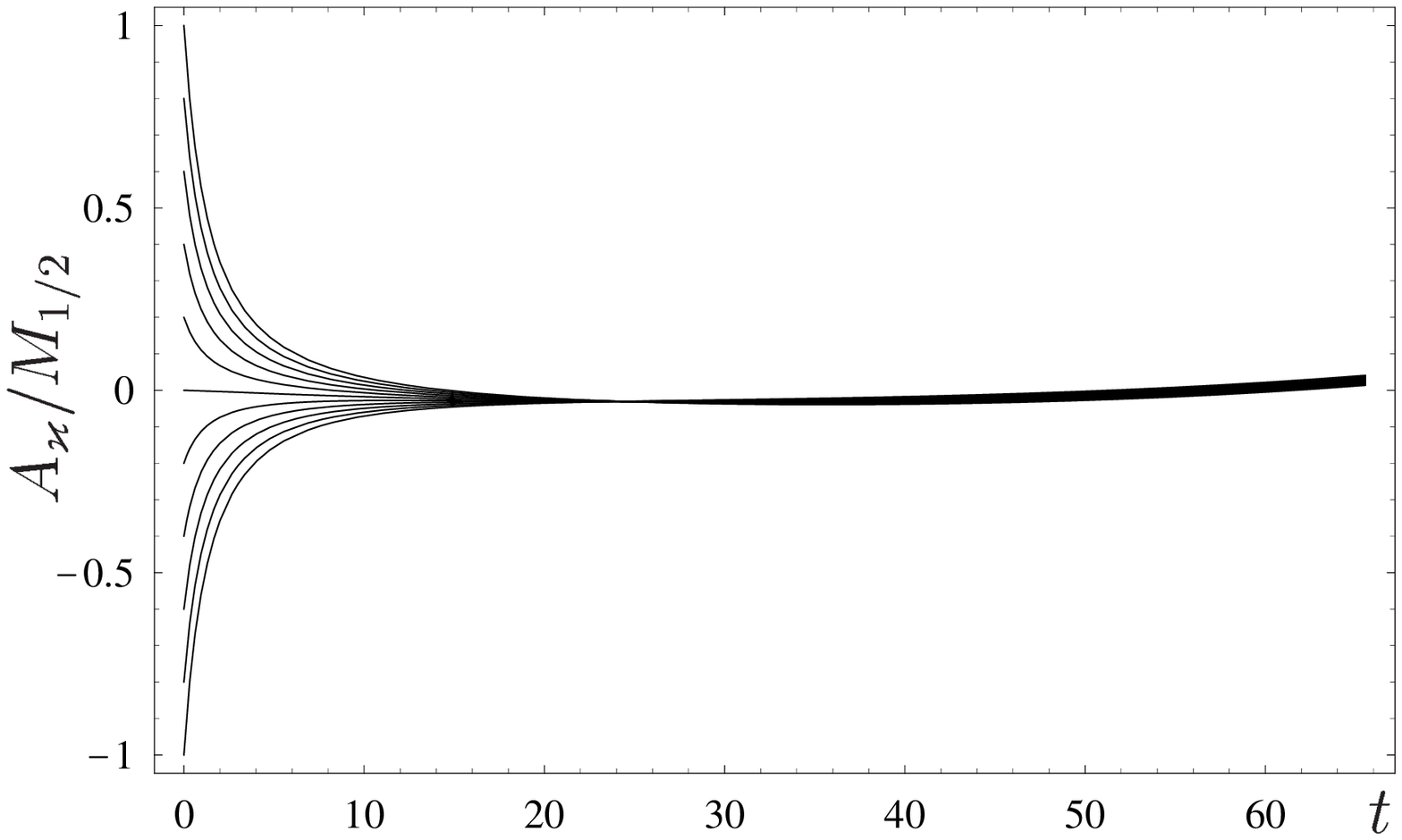}

\vspace{0mm}\hspace*{12mm}{\large\bfseries Fig.3c.}

\end{center}

\newpage

\begin{center}

\vspace*{-15mm}\includegraphics[height=69mm,keepaspectratio=true]{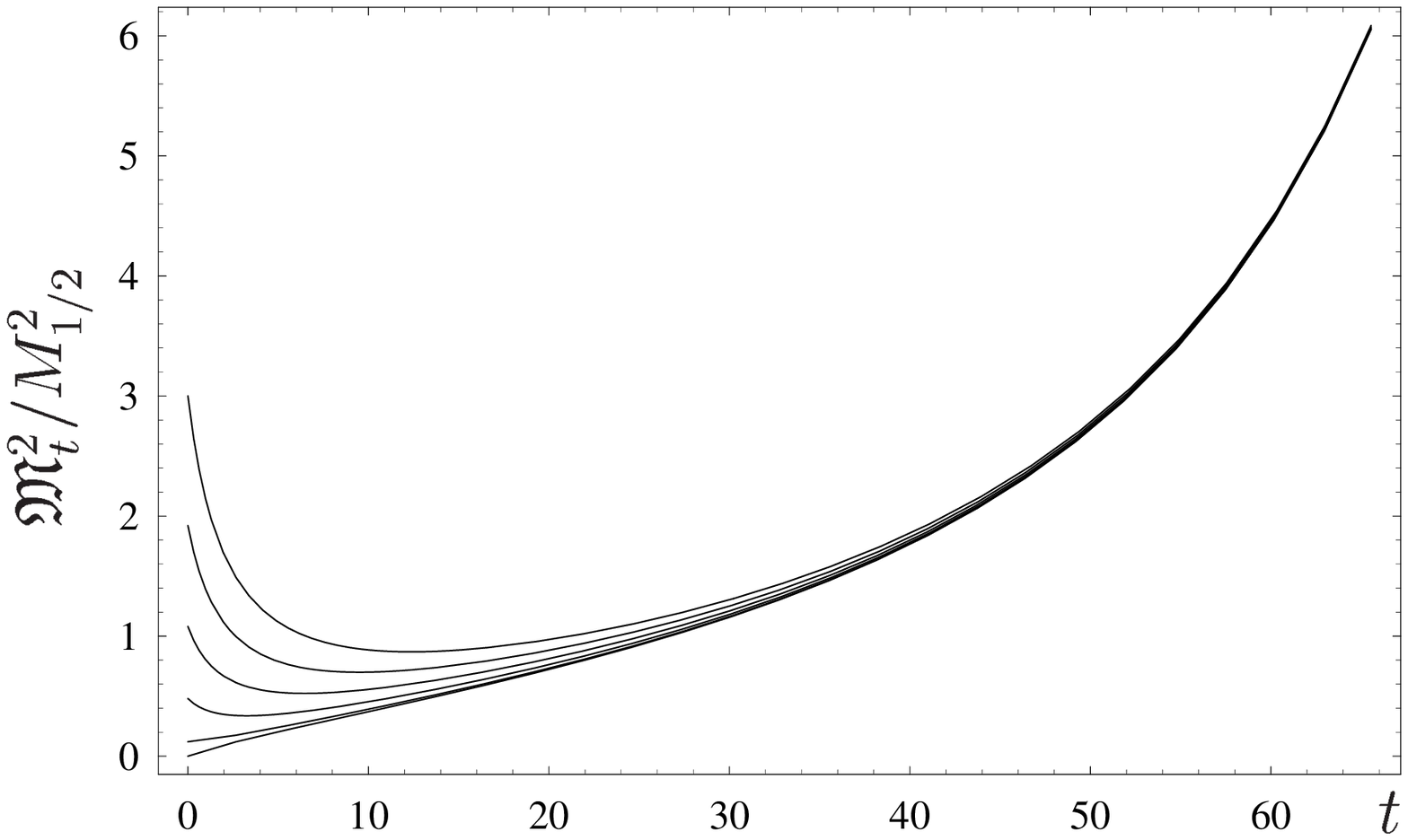}

\vspace{0mm}\hspace*{12mm}{\large\bfseries Fig.4a.}

\vspace{8mm}\includegraphics[height=69mm,keepaspectratio=true]{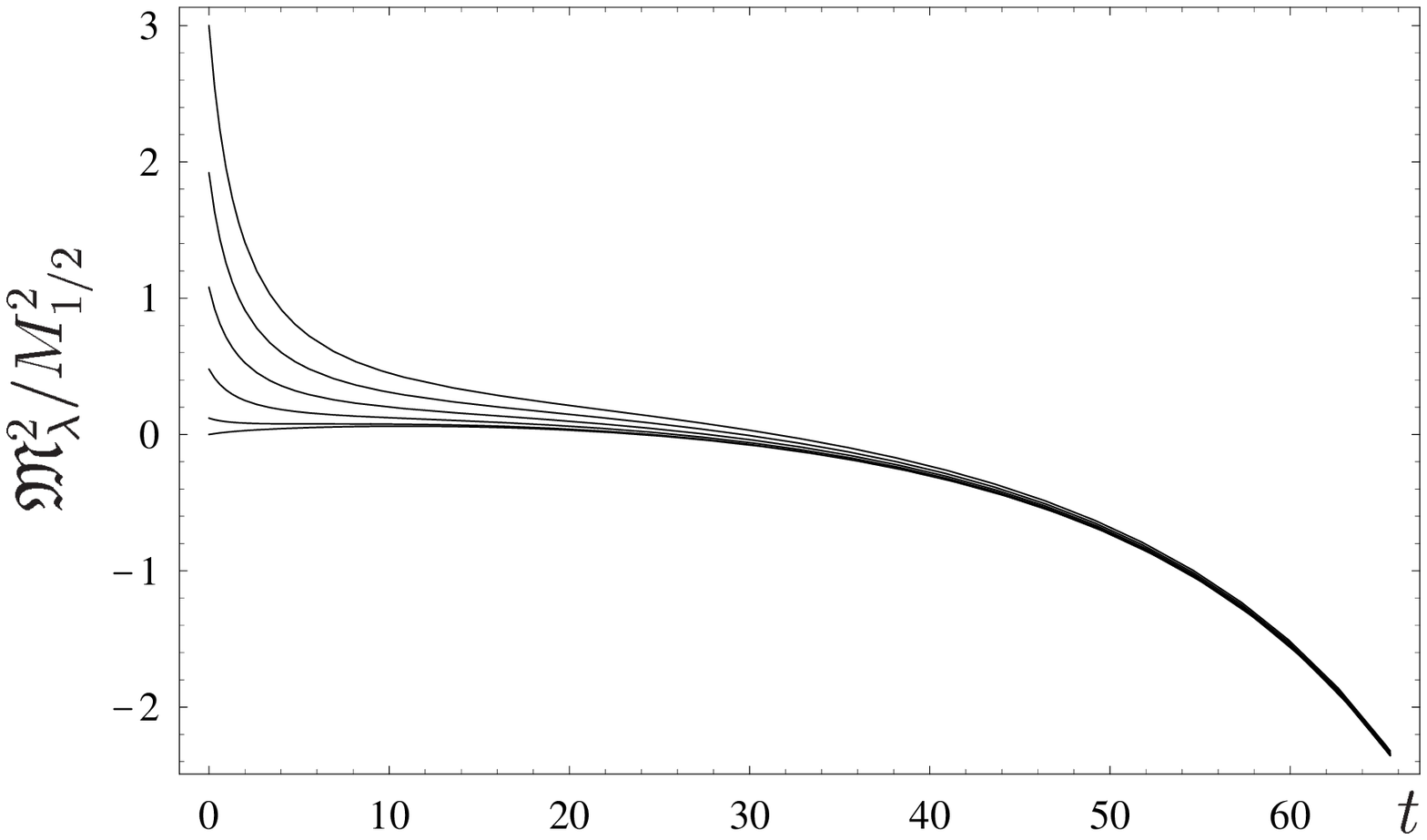}

\vspace{0mm}\hspace*{12mm}{\large\bfseries Fig.4b.}

\vspace{8mm}\includegraphics[height=69mm,keepaspectratio=true]{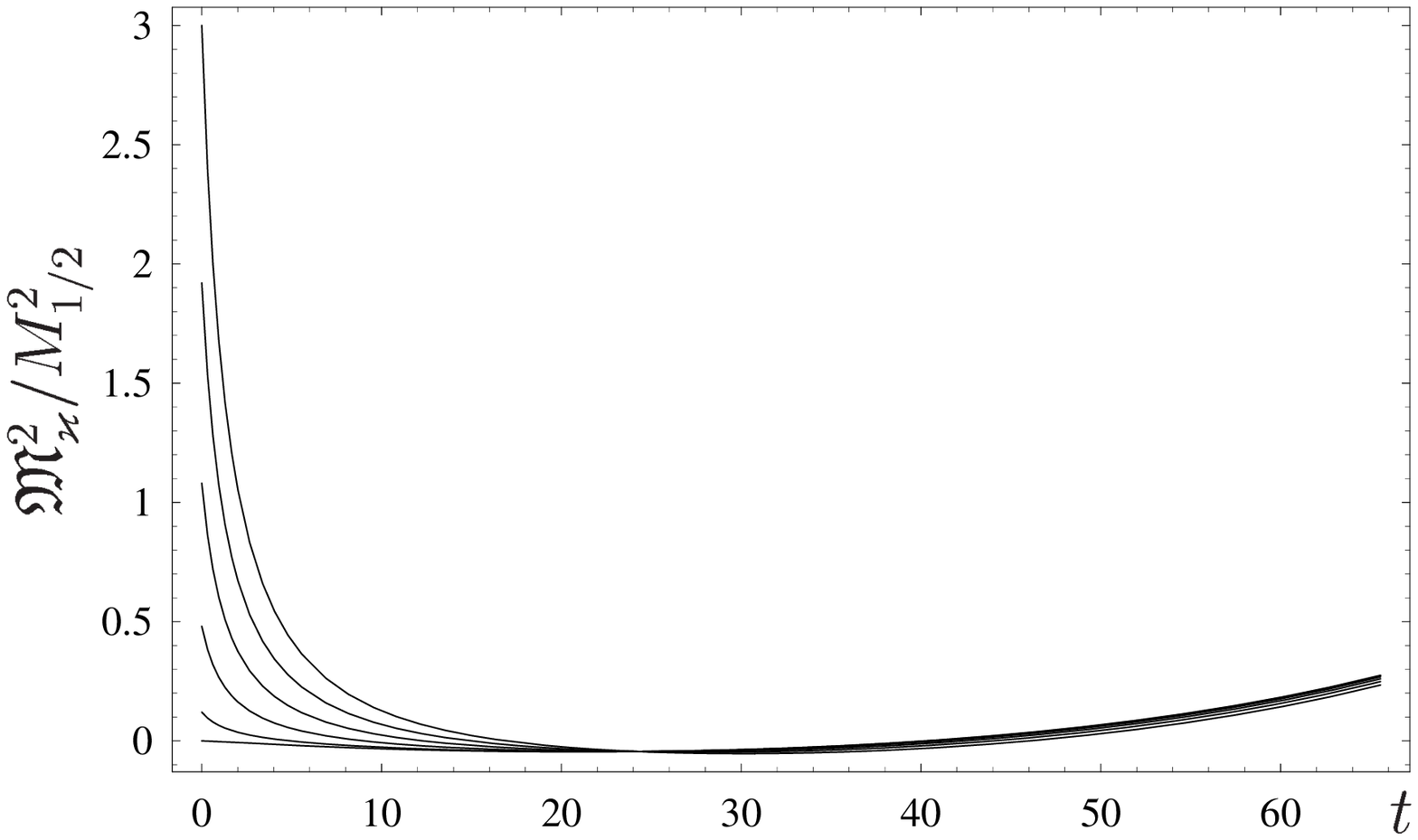}

\vspace{0mm}\hspace*{12mm}{\large\bfseries Fig.4c.}

\end{center}

\begin{landscape}

\noindent\raisebox{3mm}{\includegraphics[height=92mm,keepaspectratio=true]{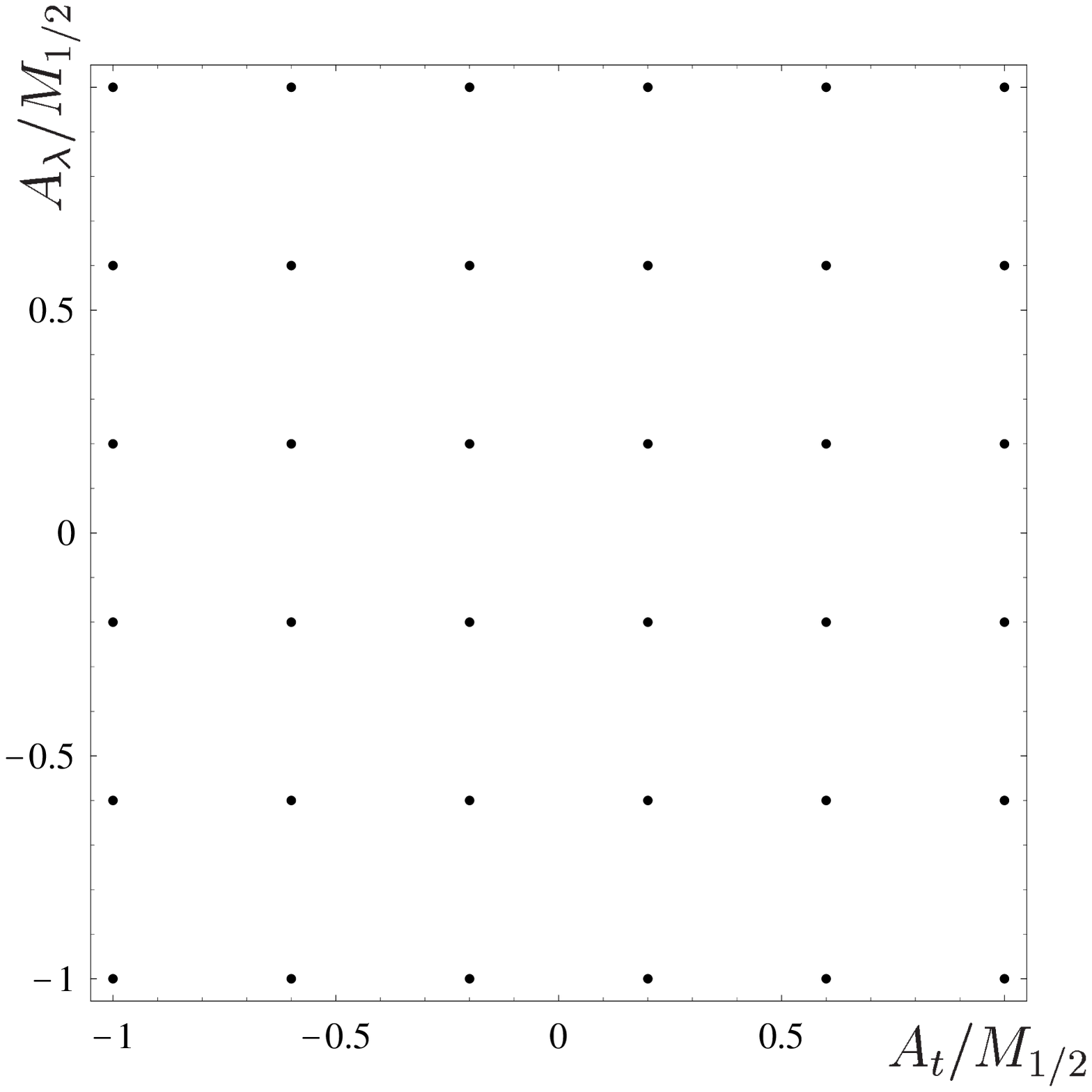}}
\hfill
\includegraphics[height=97mm,keepaspectratio=true]{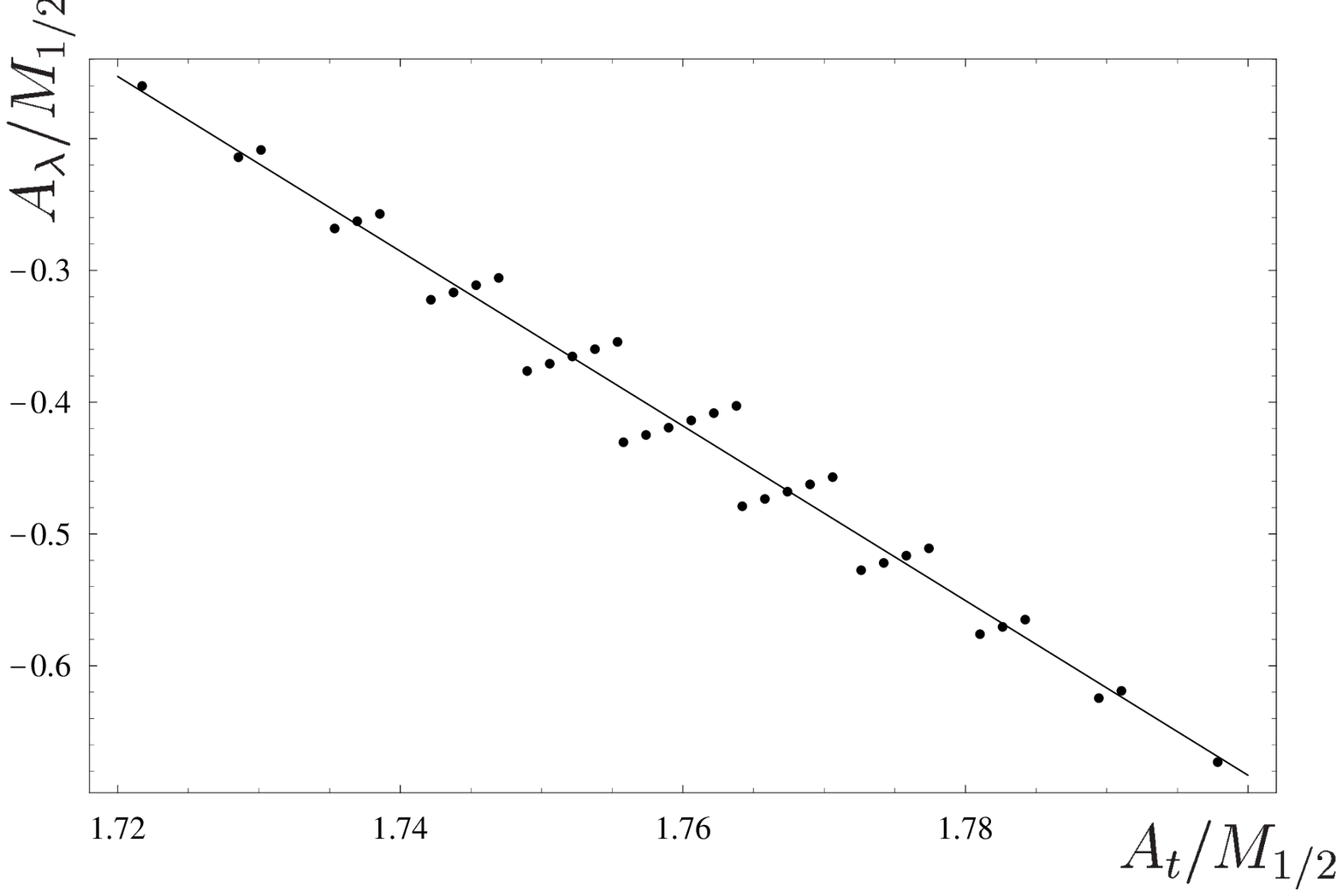}

\hspace*{32mm}{\large\bfseries
Fig.5a}\hspace*{109.5mm}{\large\bfseries Fig.5b.}

\end{landscape}

\begin{landscape}

\noindent\raisebox{2.5mm}{\includegraphics[height=92mm,keepaspectratio=true]{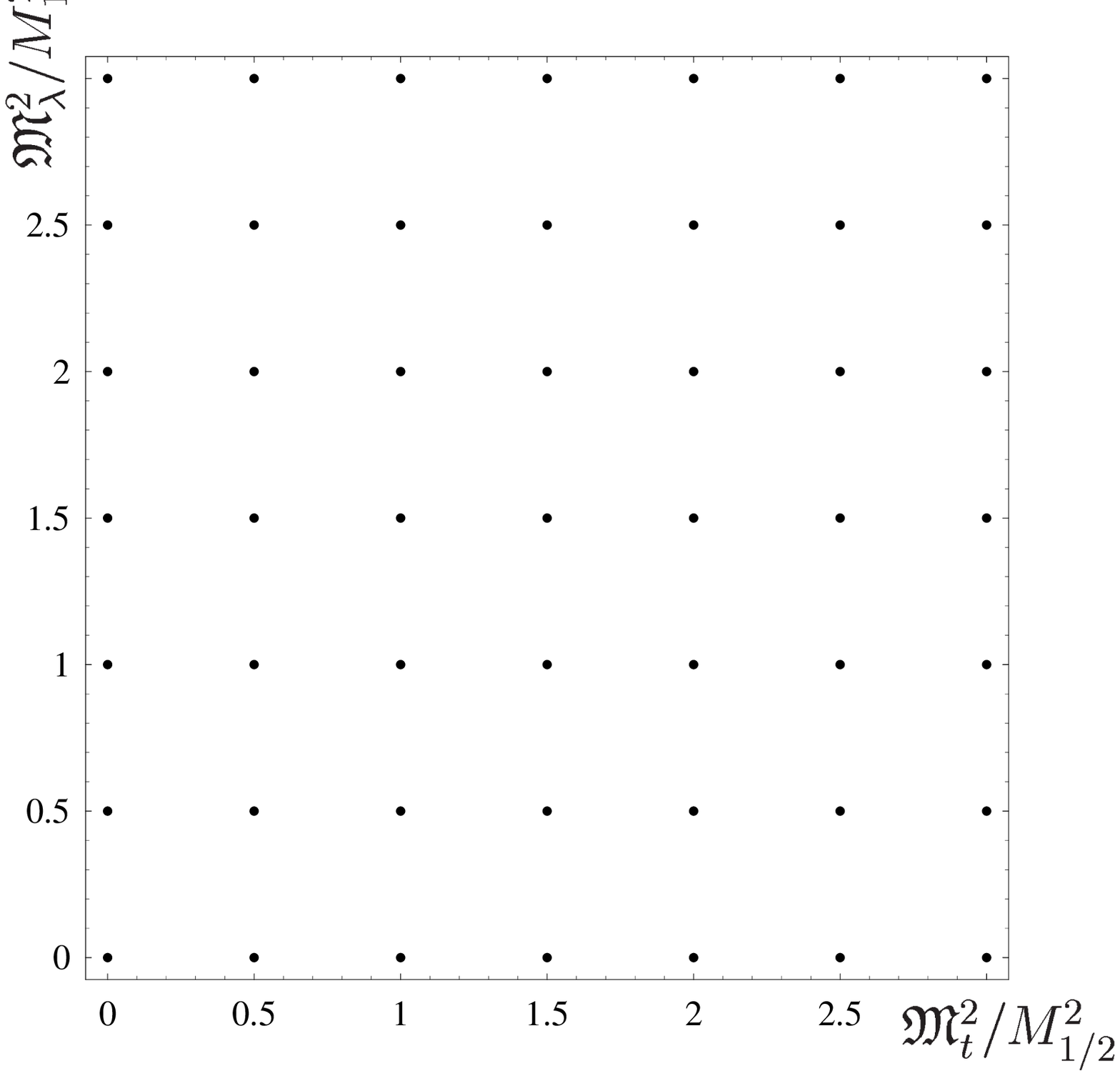}}
\hfill
\includegraphics[height=100mm,keepaspectratio=true]{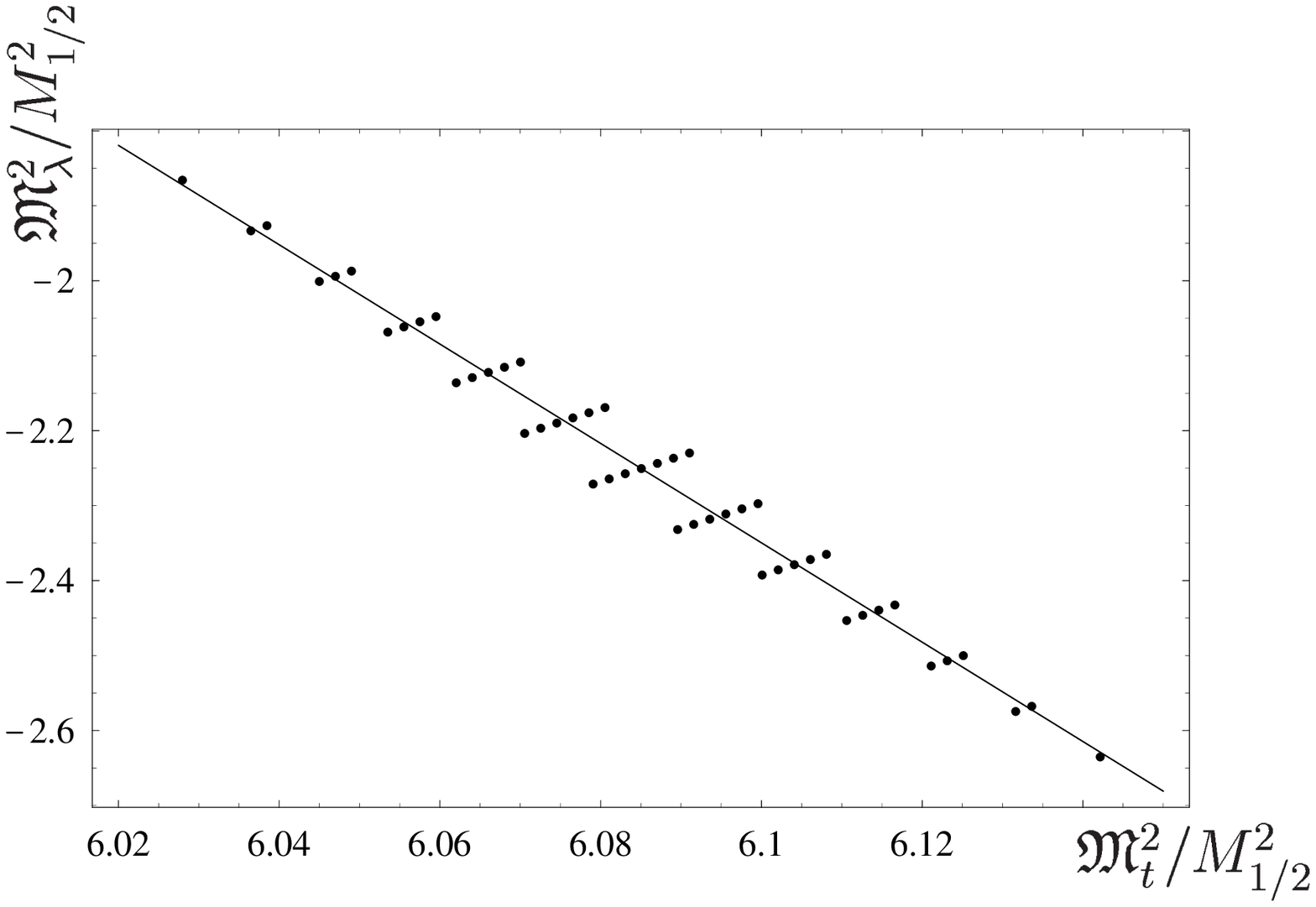}

\hspace*{30mm}{\large\bfseries
Fig.6a.}\hspace*{108.5mm}{\large\bfseries Fig.6b.}

\end{landscape}

\begin{center}

\vspace*{-15mm}\includegraphics[height=100mm,keepaspectratio=true]{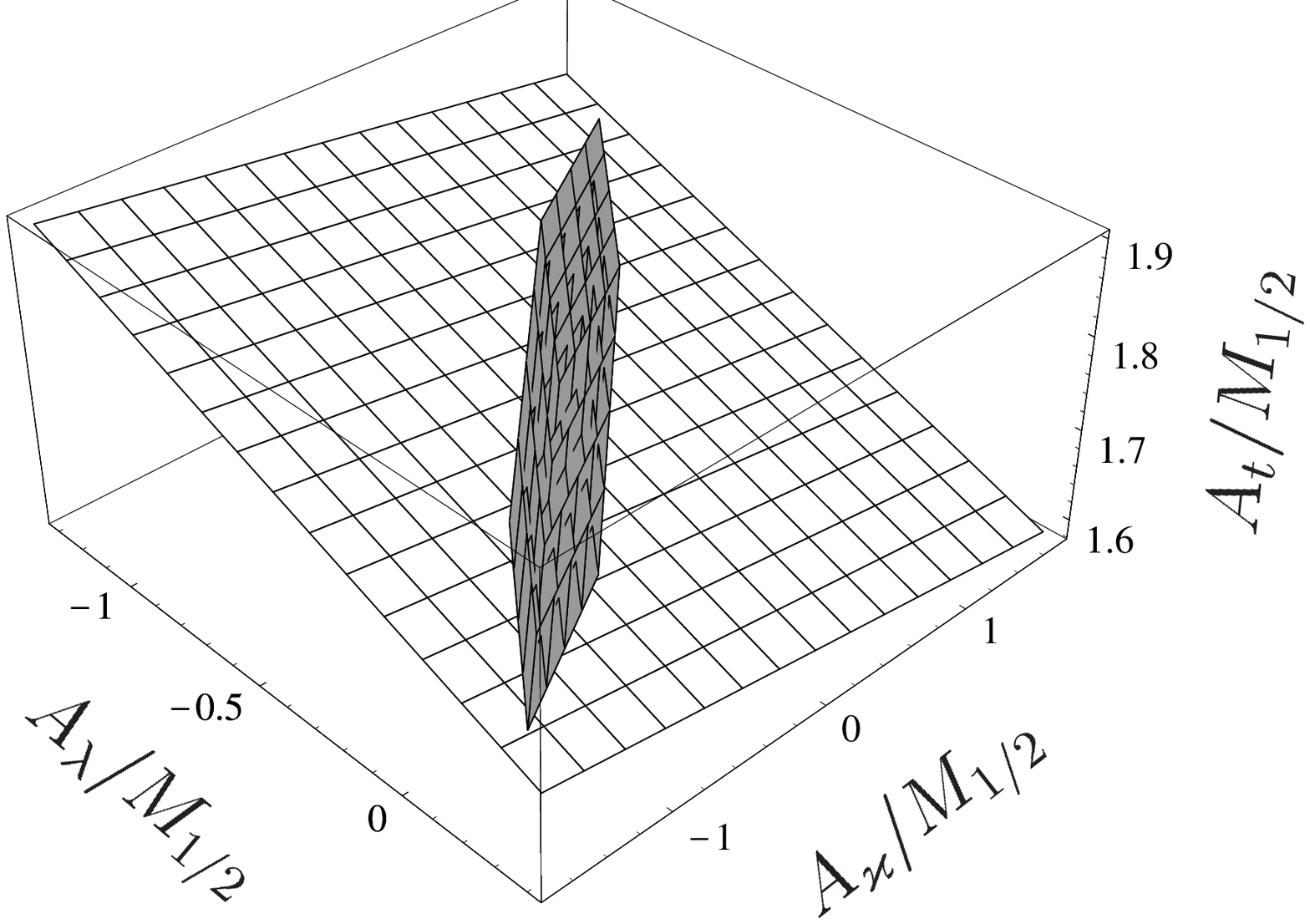}

\vspace{5mm}\hspace*{-21mm}{\large\bfseries Fig.7a.}

\vspace{15mm}\includegraphics[height=105mm,keepaspectratio=true]{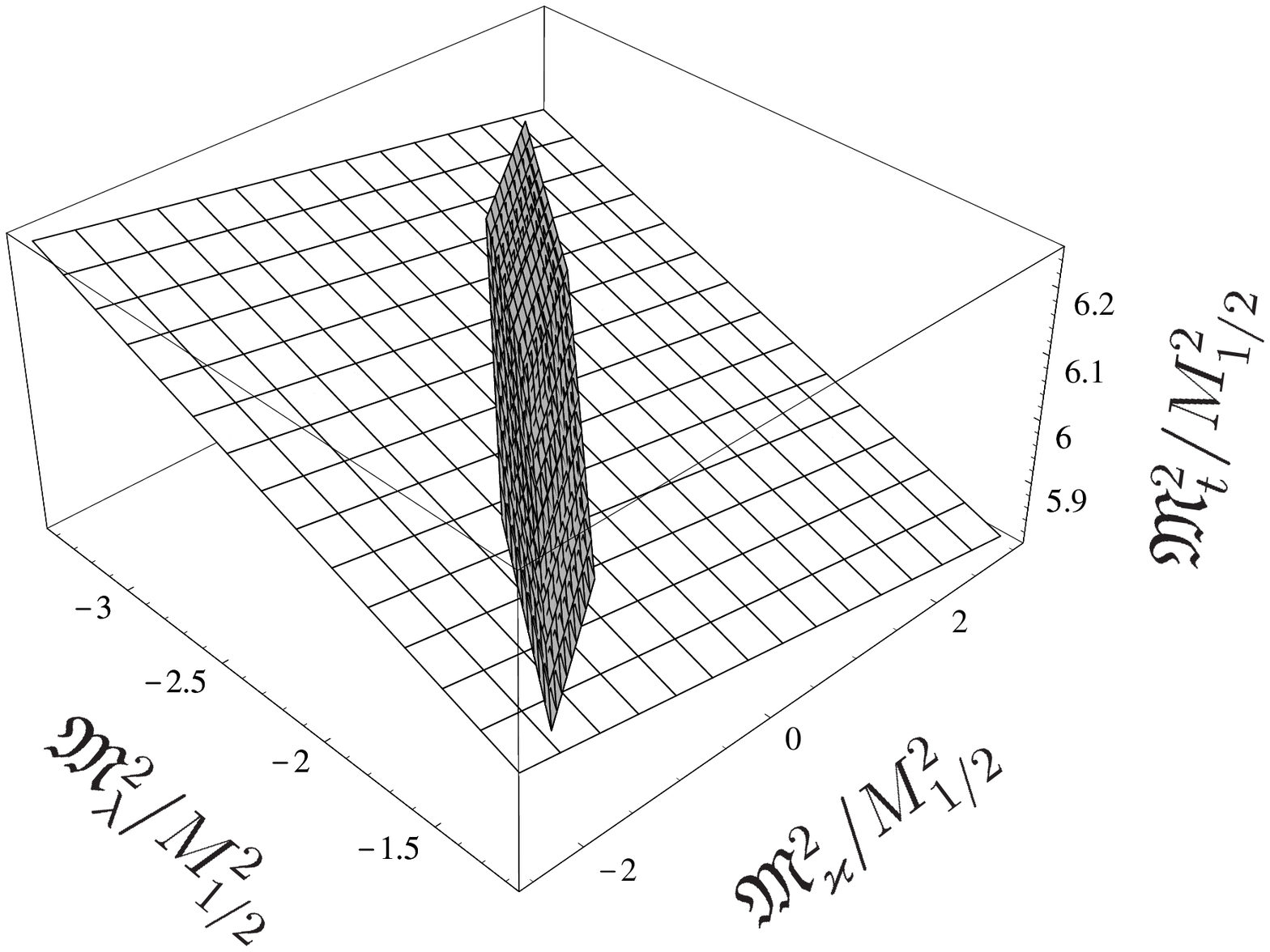}

\vspace{5mm}\hspace*{-21mm}{\large\bfseries Fig.7b.}

\end{center}

\newpage

\begin{center}

\vspace*{-15mm}\includegraphics[height=100mm,keepaspectratio=true]{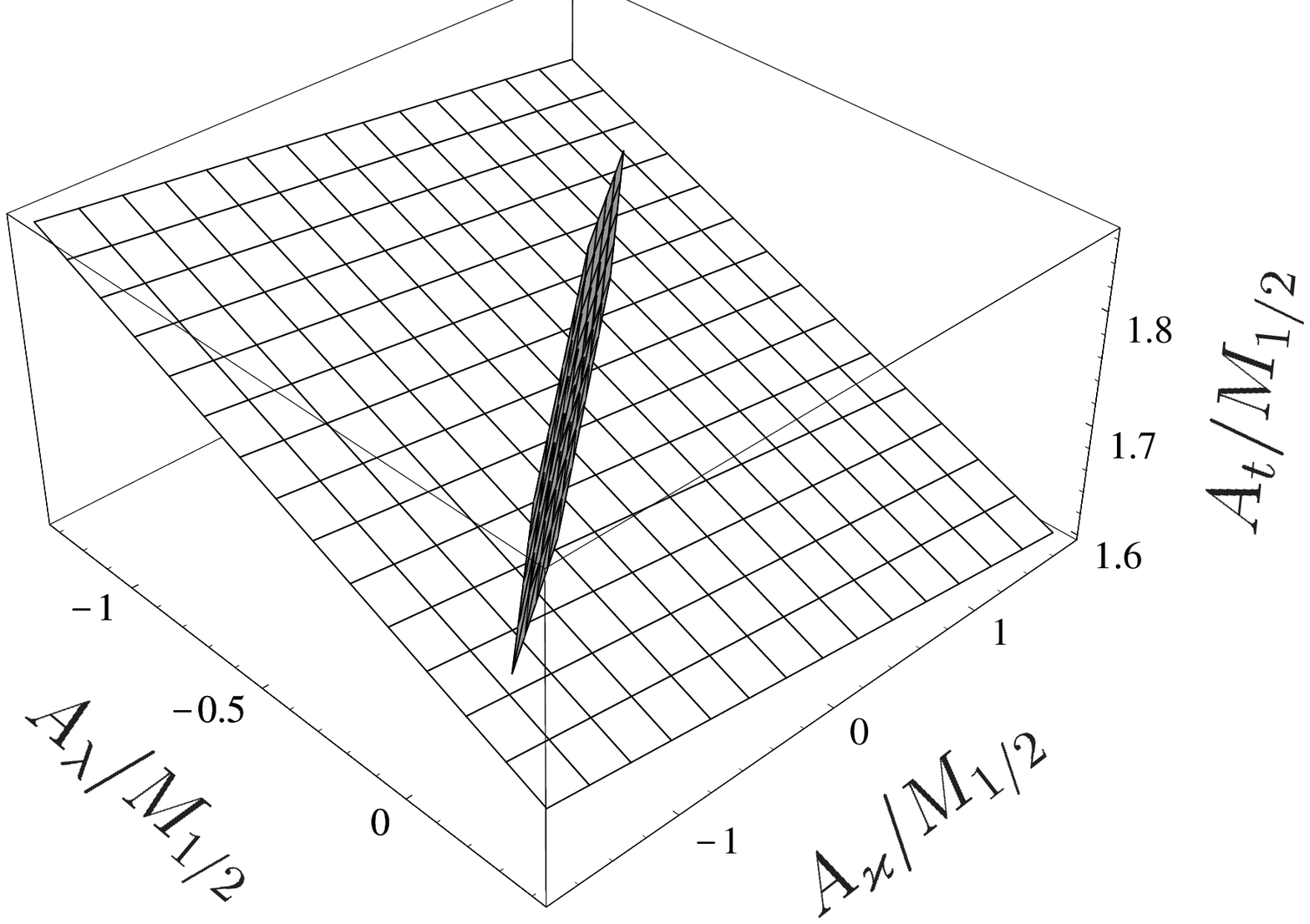}

\vspace{5mm}\hspace*{-21mm}{\large\bfseries Fig.8a.}

\vspace{15mm}\includegraphics[height=105mm,keepaspectratio=true]{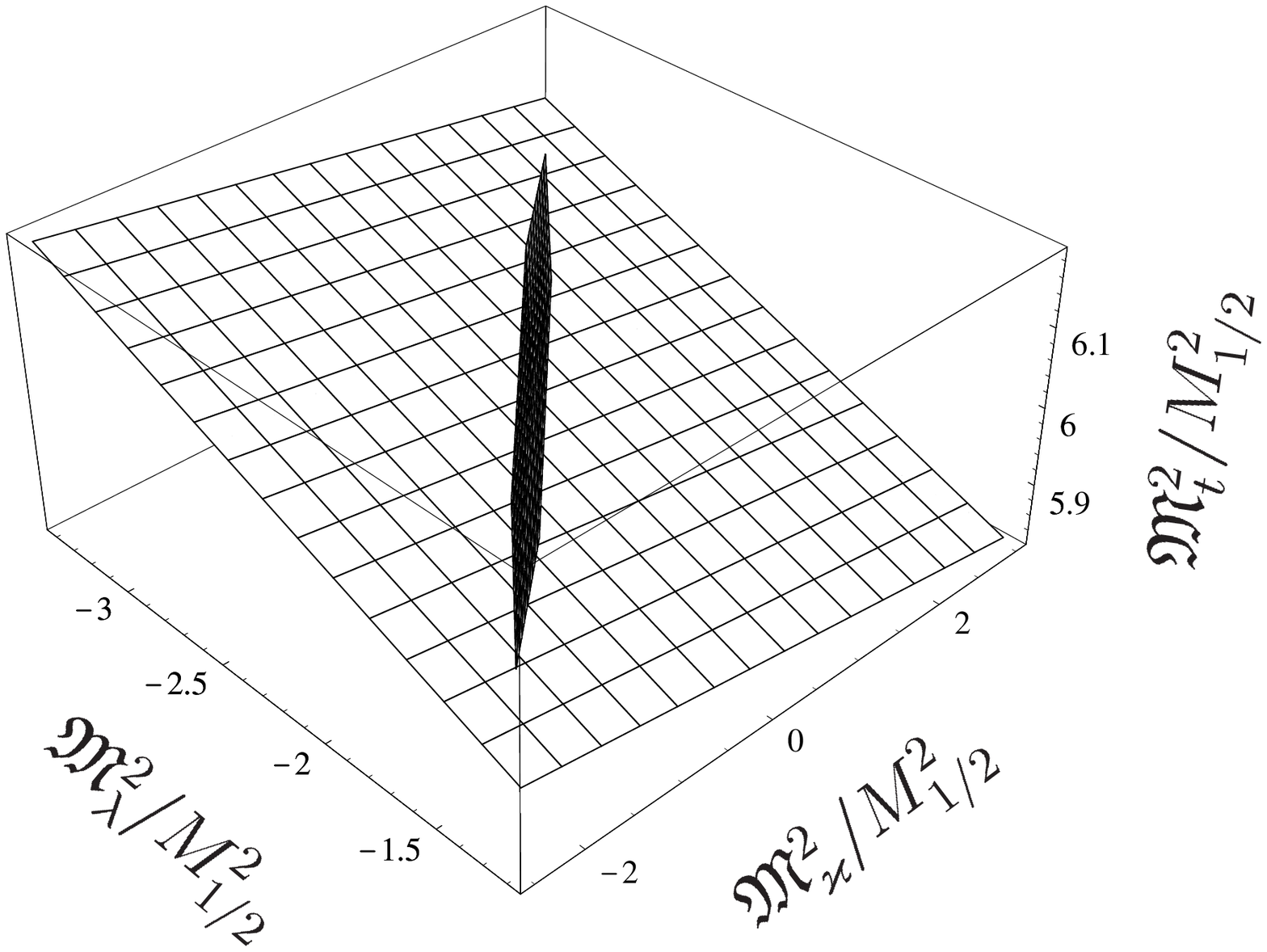}

\vspace{5mm}\hspace*{-21mm}{\large\bfseries Fig.8b.}

\end{center}

\newpage

\begin{center}

\vspace*{-15mm}\includegraphics[width=155mm,keepaspectratio=true]{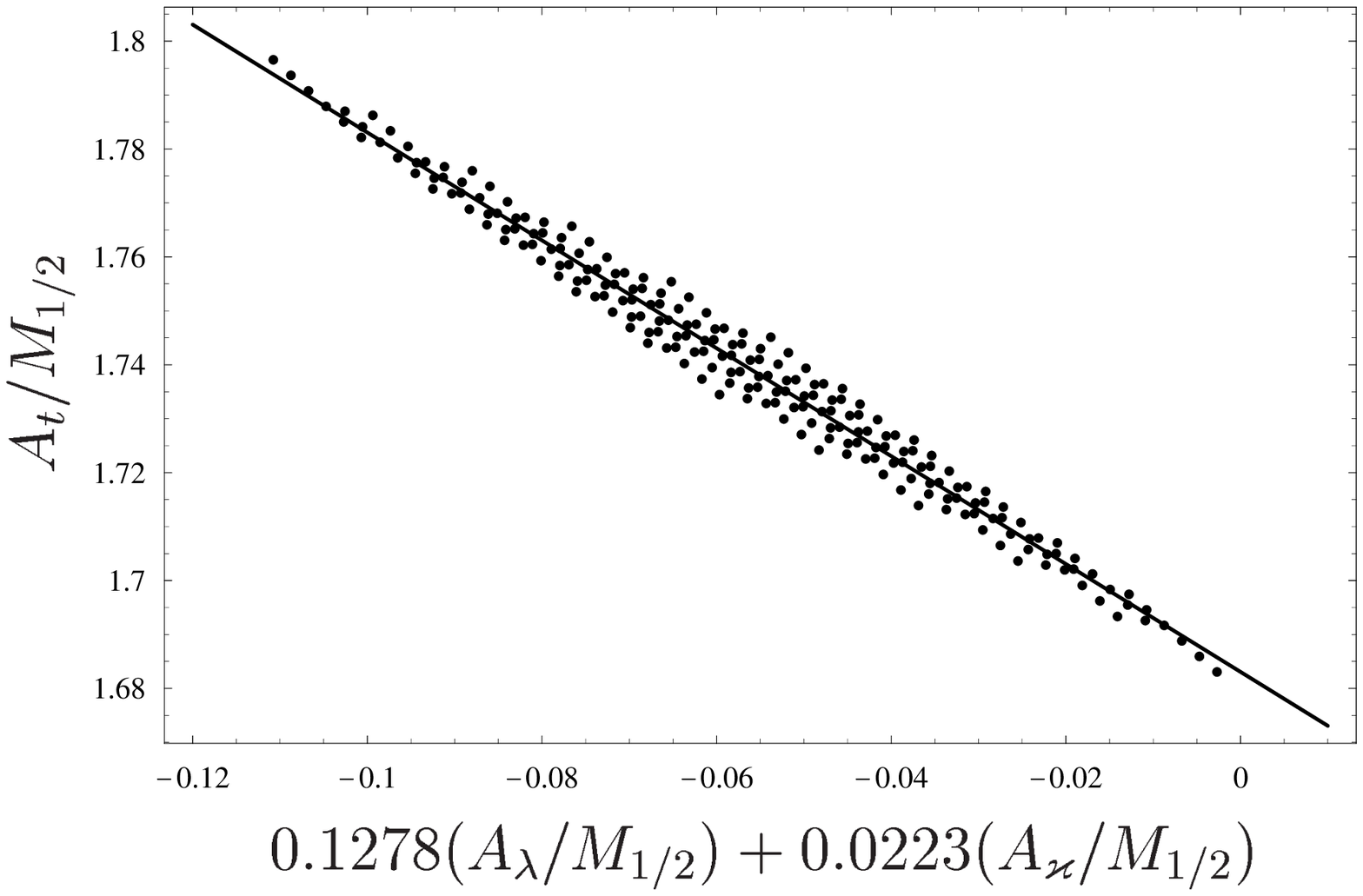}

\vspace{5mm}\hspace*{17mm}{\large\bfseries Fig.9a.}

\vspace{15mm}\includegraphics[width=155mm,keepaspectratio=true]{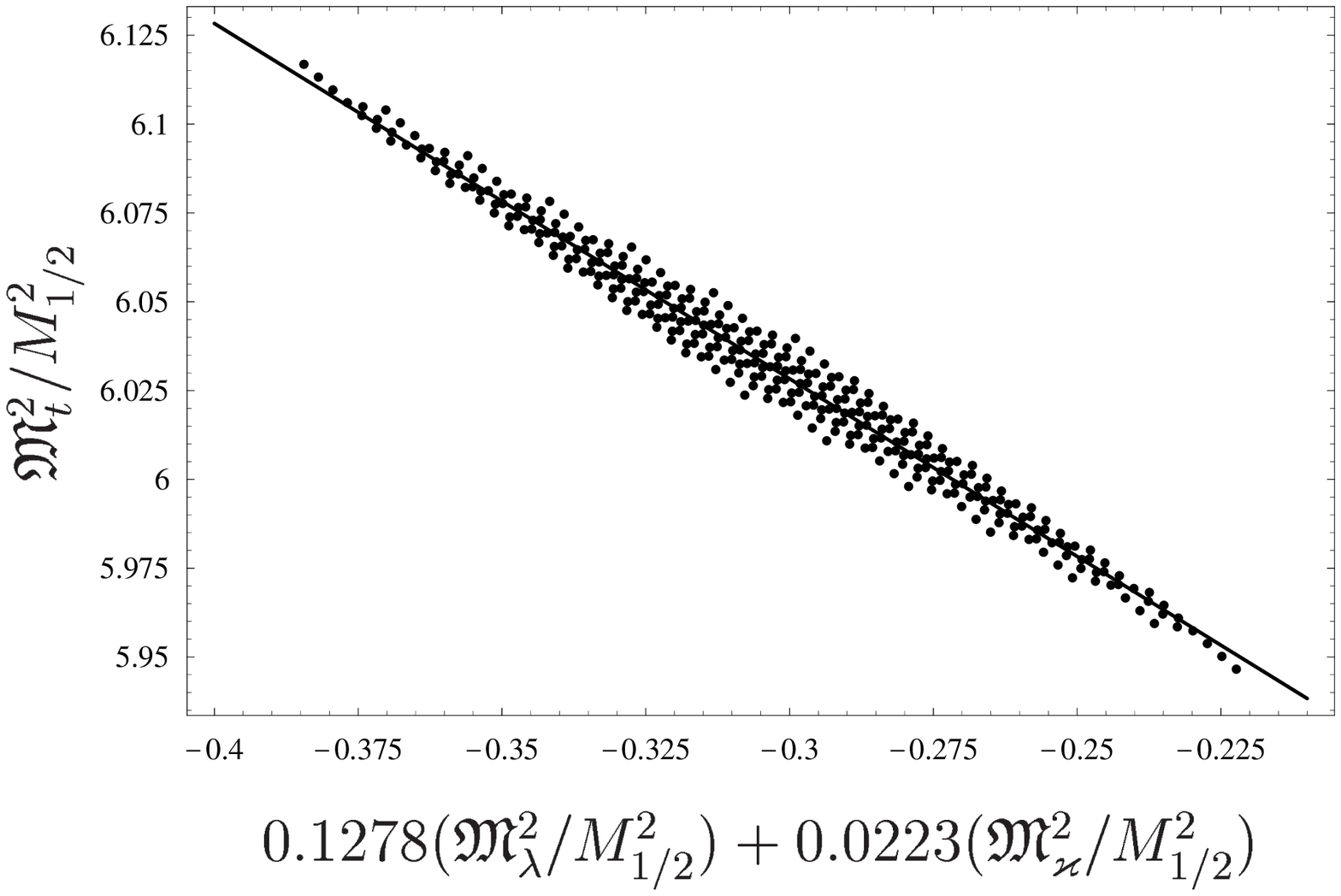}

\vspace{5mm}\hspace*{17mm}{\large\bfseries Fig.9b.}

\end{center}

\end{document}